\documentclass[11pt,a4paper,fleqn]{article}

\usepackage{report.dpo}
\newtheoremstyle{dotless}{}{}{\itshape}{}{\bfseries}{}{ }{}
\theoremstyle{dotless}

\setlist[enumerate]{itemsep=0pt}
\setlist[enumerate]{ref=\arabic*}
\usepackage{accents}
\newcolumntype{H}{>{\setbox0=\hbox\bgroup}c<{\egroup}@{}}

\DeclareFontFamily{U}{mathx}{\hyphenchar\font45}
\DeclareFontShape{U}{mathx}{m}{n}{
      <5> <6> <7> <8> <9> <10>
      <10.95> <12> <14.4> <17.28> <20.74> <24.88>
      mathx10
      }{}
\DeclareSymbolFont{mathx}{U}{mathx}{m}{n}
\DeclareFontSubstitution{U}{mathx}{m}{n}
\DeclareMathAccent{\widecheck}{0}{mathx}{"71}
\DeclareMathAccent{\wideparen}{0}{mathx}{"75}
\DeclareMathAccent{\widehatown}{0}{mathx}{"70}

\newcommand*\xOverBar[1]{%
   \hbox{%
     \vbox{%
       \hrule height 0.4pt 
       \kern0.3ex
       \hbox{%
         \kern-0.1em
         \ensuremath{#1}%
         \kern-0.1em
       }%
     }%
   }%
} 

\newcommand*\xBla[1]{%
   \hbox{%
     \vbox{%
       \hrule height 0.4pt 
       \kern0.3ex
        \hrule height 0.4pt 
        \kern0.3ex%
       \hbox{%
         \kern-0.1em
         \ensuremath{#1}%
         \kern-0.1em
       }%
     }%
   }%
} 

\newcommand*\xUnderBar[1]{%
   \hbox{%
     \vbox{%
       \hrule height 0.4pt 
       \kern-1.3ex
       \hbox{%
         \kern -0.3ex
         \ensuremath{#1}%
         \kern-0.1em
       }%
     }%
   }%
}

\newcommand*\xUnderOverBar[1]{%
   \hbox{%
     \vbox{%
       \hrule height 0.8pt 
       \kern-1.6ex
       
        \hrule height 0.4pt 
       \kern+0.3ex
       \hbox{%
         \kern -0.3ex
         \ensuremath{#1}%
         \kern-0.1em
       }%
     }%
   }%
}

\newcommand*\xTwoUnderOverBar[1]{%
   \hbox{%
     \vbox{%
     
      \hrule height 0.4pt 
       \kern-0.3ex
       
       \hrule height 0.4pt 
       \kern-1.9ex
       
        \hrule height 0.4pt 
       \kern+0.3ex
       
         \hrule height 0.4pt 
       \kern+0.3ex
       
       \hbox{%
         \kern -0.3ex
         \ensuremath{#1}%
         \kern-0.1em
       }%
     }%
   }%
} 

\newcommand{\xTwoTop}[1]{\accentset{=}{x}_{#1}}
\newcommand{\costTwoTop}[1]{\accentset{=}{c}_{#1}}

\newcommand{\xTwoBottom}[1]{\underaccent{=}{x}_{#1}}
\newcommand{\costTwoBottom}[1]{\underaccent{=}{c}_{#1}}

\newcommand{\xTopBottom}[1]{\underaccent{\bar}{\bar{x}}_{#1}}
\newcommand{\costTopBottom}[1]{\underaccent{\bar}{\bar{c}}_{#1}}

\newcommand{\xPickerBottomRight}[1]{\underaccent{\rightarrow}{{x}}_{#1}^{p}}
\newcommand{\xPickerTopRight}[1]{\accentset{\rightarrow}{x}_{#1}^{p}}

\newcommand{\xPickerBottomLeft}[1]{\underaccent{\leftarrow}{{x}}_{#1}^{p}}
\newcommand{\xPickerTopLeft}[1]{\accentset{\leftarrow}{x}_{#1}^{p}}

\newcommand{\xTwoTopBottom}[1]{\underaccent{=}{\accentset{=}{x}}_{#1}}
\newcommand{\costTwoTopBottom}[1]{\underaccent{=}{\accentset{=}{c}}_{#1}}

\newcommand{\xLast}[1]{\tilde{x}_{#1}}

\newcommand{\xUp}[1]{x^{\shortmid}_{#1}}
\newcommand{\costUp}[1]{c^{\shortmid}_{#1}}
\newcommand{\xTwoUp}[1]{x^{\shortmid\shortmid}_{#1}}
\newcommand{\costTwoUp}[1]{c^{\shortmid\shortmid}_{#1}}

\newcommand{\costPickBottom}[2] {\underaccent{\widehatown}{c}_{#1#2}}
\newcommand{\xPickBottom}[2]{\underaccent{\widehatown}{x}_{#1#2}}  
\newcommand{\costPickTop}[2]{\widecheck{c}_{#1#2}}     
\newcommand{\xPickTop}[2]{\widecheck{x}_{#1#2}}

\newcommand{\itemsBottom}[2] {\underaccent{\widehatown}{u}_{#1#2}}
\newcommand{\itemsTop}[2]{\widecheck{u}_{#1#2}}  

\newcommand{\xPickTopBottom}[1]{\widehatown{\underaccent{\widecheck}{x}}_{#1}}
\newcommand{\xComponent}[1]{\tau_{#1}}

\newcommand{\xPick}[2]{x_{#1#2}}

\newcommand{\capBottom}[1]{\underaccent{\circ}{\lambda}_{#1}}
\newcommand{\capTop}[1]{\accentset{\circ}{\lambda}_{#1}}

\newcommand{\depotTop}[1]{\accentset{\circ}{\theta}_{#1}}

\newcommand{\degreeEvenTop}[1]{\accentset{\circ}{\pi}_{#1}}
\newcommand{\degreeEvenBottom}[1]{\underaccent{\circ}{\pi}_{#1}}

\newcommand{\degreeEvenTopReturn}[1]{\accentset{\circ}{\rho}_{#1}}
\newcommand{\degreeEvenBottomReturn}[1]{\underaccent{\circ}{\rho}_{#1}}

\newcommand{\yUp}[1]{y^{\shortmid}_{#1}}
\newcommand{\yTop}[1]{\bar{y}_{#1}}
\newcommand{\yTopBottom}[1]{\underaccent{\bar}{\bar{y}}_{#1}}
\newcommand{\yBottom}[1]{\underaccent{\bar}{y}_{#1}}
\newcommand{\openDepotTop}[1]{\accentset{\circ}{\omega}_{#1}}
\newcommand{\openDepotBottom}[1]{\underaccent{\circ}{\omega}_{#1}}

\newcommand{\aisle}{j}
\newcommand{\previousAisle}{j-1}
\newcommand{\nextAisle}{j+1}
\newcommand{\article}{i}
\newcommand{\depotAisle}{l}

\newcommand{\cost}[1]{c_{#1}}

\newcommand{\numberOfAisles}{m}
\newcommand{\numberOfCells}{n}

\newcommand{\myArticle}{h}
\newcommand{\setOfArticles}{\mathcal{H}}
\newcommand{\setOfCellsInAisleForArticle}[2]{\mathcal{I}_{#1#2}}
\newcommand{\setOfArticlesInAisle}[1]{\mathcal{I}_{#1}}
\newcommand{\setOfAisles}{\mathcal{J}}

\newcommand{\firstAisle}{0}
\newcommand{\lastAisle}{\numberOfAisles-1}
\newcommand{\lastAisleVis}{\numberOfAisles-1}

\newcommand{\demand}[1]{b_{#1}}
\newcommand{\demandTwo}[3]{b_{#1#2#3}}
\newcommand{\capacity}[2]{q_{#1#2}}
\newcommand{\capacityTwo}[3]{q_{#1#2#3}}

\newcommand{\runtimeSup}[2]{$t_{#1}^{#2}$}
\newcommand{\runtime}[1]{$t_{#1}$}
\newcommand{\solvedInst}{opt}
\newcommand{\numberOfArticles}{a}

\newcommand{\vertical}{vertical\xspace}
\newcommand{\horizontal}{horizontal\xspace}

\usepackage[authormarkuptext=name]{changes}

\definechangesauthor[name={}, color=red]{michael}
\definechangesauthor[name={}, color=blue]{max}

\allowdisplaybreaks
\begin{document}

\title{Modeling Single Picker Routing Problems in Classical and Modern Warehouses}

\category{Working Paper DPO-2018-11 (version 1, 04.11.2018)}

\authors{\textbf{Dominik Goeke and Michael Schneider}\\
goeke$|$schneider@dpo.rwth-aachen.de\\
Deutsche Post Chair\,--\,Optimization of Distribution Networks\\
RWTH Aachen University\\[2ex]}

\abstract{The standard single picker routing  problem (SPRP) seeks the cost-minimal tour to collect a set of given articles in a rectangular single-block warehouse with parallel picking aisles and a dedicated storage policy, i.e, each SKU is only available from one storage location in the warehouse. We present a compact formulation that forgoes classical subtour elimination constraints by directly exploiting two of the properties of an optimal picking tour used in the dynamic programming algorithm of Ratliff and Rosenthal~(1983). We extend the formulation to three important settings prevalent in modern e-commerce warehouses: scattered storage, decoupling of picker and cart, and multiple end depots. In numerical studies, our formulation outperforms existing standard SPRP formulations from the literature and proves able to solve large instances within short runtimes. Realistically sized instances of the three problem extensions can also be solved with low computational effort. We find that decoupling of picker and cart can lead to substantial cost savings depending on the speed and capacity of the picker when traveling alone, whereas additional end depots have rather limited benefits in a single-block warehouse.\\[1ex]
\textbf{Keywords:} \textit{warehouse management, picker routing, scattered storage, decoupling, multiple end depots}}

\logo{\includegraphics[scale=0.75]{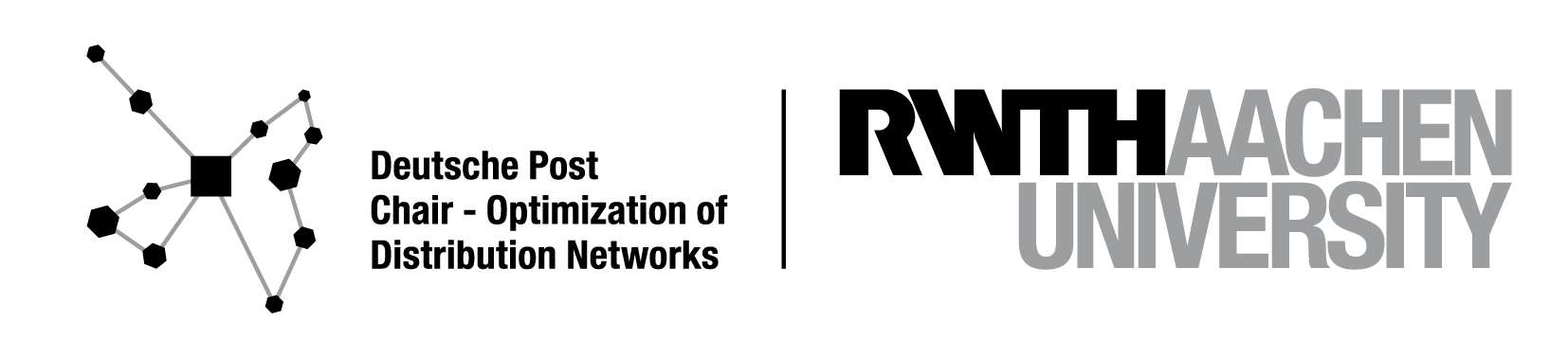}}
\titlepage

\section{Introduction}
\label{sec:intro}
Order picking is a central and labor-intensive task in warehouses. The aim of {single picker routing problems} (SPRPs) is to determine a picker tour of minimum cost---starting from and ending at a depot---to collect all stock keeping units (SKUs) contained in a pick list from their storage locations in the warehouse. The cost of a tour is typically measured as distance or time. Single-block SPRPs are defined on a rectangular warehouse, in which the SKUs are stored in racks along both sides of multiple parallel picking aisles that are enclosed by a storage-free cross aisle at the top and at the bottom (see Figure~\ref{fig:warehouse}). Each of the picking aisles contains a number of picking positions, and multiple different SKUs can be located at the same picking position. In single-block SPRPs, we do not distinguish between a picking request from the rack on the left side, on the right side, or from both sides of a picking position. All these cases are treated equally, and only the travel cost to reach the picking positions in the aisles is of relevance. Therefore, a pick list translates into a set of required picking positions that the picker needs to visit. 

The single-block SPRP with dedicated storage, in which each SKU is only available from one picking position in the warehouse, is the most well-studied SPRP variant, and is denoted as standard SPRP in the following. In a seminal work, \citet{Ratliff:1983} introduce a dynamic programming (DP) algorithm to solve the standard SPRP to optimality  with  a runtime linear in the number of picking aisles. \cite{Roodbergen:2001} extend the DP to two-block warehouses, and \citet{Pansart:2018} present a DP that is applicable to warehouses with an arbitrary number of blocks, however, the runtime complexity is exponential in the number of cross aisles. 

SPRPs can also be tackled using mathematical formulations that are solved with the help of optimization software. To address the standard SPRP,
\citet{Scholz:2016} reduce the number of vertices that have to be considered in each picking aisle based on the fact that the largest gap in an aisle is never traversed if the aisle is entered from top and bottom, originally discussed in \citet{Ratliff:1983}.  On the resulting graph, they solve a single-commodity flow formulation of a traveling salesman problem (TSP) variant that contains optional vertices indicating the direction of travel at the entry and exit of each picking aisle. In this way, they obtain a formulation whose size is linear to the number of picking aisles. This formulation is compared to three TSP formulations defined on a complete graph spanning the SKUs to be picked and one Steiner TSP formulation that is adapted to single-block warehouses. The authors demonstrate on a large set of test instances that their formulation is superior with regard to the size of the instances that can be solved and the runtimes for solving the instances. Their formulation can also be extended to multi-block warehouses, but they only present results for the standard SPRP. \citet{Pansart:2018} present a model of the SPRP in multi-block warehouses that is based on a single-commodity flow formulation of the Steiner TSP. The authors use a procedure similar to the one described in \citet{Scholz:2016} to reduce the number of vertices, and the number of arcs is decreased by solving the minimum 1-spanner problem using a commercial solver. In addition, valid inequalities exploiting the special structure of the warehouse are added, and the solver is provided with upper bounds that are computed using a freely available version of the heuristic of \citet{Lin:1973}. On single-block warehouse instances, their formulation is clearly superior to the formulation of \citet{Scholz:2016}. 

We propose a compact formulation of the standard SPRP that directly exploits two properties of an optimal tour used in the algorithm of \citet{Ratliff:1983}: (i)~two consecutive picking aisles can only be connected using four possible configurations, and (ii)~to prevent the generation of isolated subtours, it is sufficient to ensure that the tour is always connected and the degree of the connections at the top and bottom of each picking aisle is of even degree. Thus, no classical subtour elimination constraints are needed. Although we do not rely on preprocessing or the addition of cuts to speed up the solution of our model, our formulation vastly outperforms the one of \citet{Scholz:2016} and is approximately six times faster than the one of \citet{Pansart:2018} on a set of benchmark instances with up to 30 picking aisles and 45 required picking positions using a comparable computer. Our approach shows a convincing scaling behavior and is able to solve instances with 1000 aisles, 1000 available picking positions per aisle, and 1000 required picking positions in approximately two minutes.

In addition, our model can be extended to cope with three important settings relevant to modern e-commerce warehouses: 
\begin{itemize}
\item \textbf{Scattered storage:} In warehouses with scattered storage, any SKU can be available from more than one picking position. This setting plays a major role in modern e-commerce warehouses of companies like Amazon or Zalando and is receiving growing attention from the scientific community \citep[see, e.g.,][]{Boysen:2018, Weidinger:2018b}. \citet{Daniels:1998} propose a TSP formulation for the SPRP with scattered storage for arbitrary warehouse layouts and compare several heuristics. \citet{Weidinger:2018a} shows that the single-block SPRP with scattered storage is NP-hard. He proposes a heuristic based on the decomposition of the problem into a selection and a routing problem. As comparison method, the formulation of \citet{Daniels:1998} using \citet*{Miller:1960} subtour elimination constraints is realized with Gurobi. Given a time limit of three hours, the formulation is able to solve most of the single-block warehouse instances generated by the authors with three picking aisles, 30 picking positions per aisle, and pick lists with up to seven requested SKUs. In contrast, the extension of our formulation to the single-block SPRP with scattered storage solves large instances with up to 100 picking aisles, 180 picking positions per aisle, and pick lists containing up to 30 SKUs within short runtimes of at most three minutes.
\item \textbf{Decoupling of picker and cart:} In manual order picking, items are typically retrieved from the warehouse by a picker pushing a cart, so that multiple items can be picked during one tour. To speed up the order picking, Zalando, a large fashion online retailer, allows pickers to park the cart during the tour, retrieve a few items traveling on their own, then return to the cart and continue their tour (comparable to the picking behavior of people in supermarkets). The company also incorporates this option when planning picker tours \citep{Zalando:2014,Nvidia:2015}; however, no mathematical model or algorithm has yet been published.
We extend our formulation to the single-block SPRP with decoupling of picker and cart and investigate the potential time savings of this approach depending on the carrying capacity and the speed of the picker without cart.
\item \textbf{Multiple end depots:} To reduce unnecessary trips back to a central depot, warehouse managers can use multiple end depots at which collected items can be dropped off, e.g., at dedicated positions of a conveyor belt. \Citet{DeKoster:1998} consider the single-block SPRP with decentralized depositing, in which they assume that it is possible to drop items anywhere along the upper or lower cross aisle. \citet{Scholz:2016} show how to extend their formulation to this problem variant, but they only present results for the single depot case. We extend our formulation to single-block SPRP with multiple end depots, and we investigate the potential cost savings depending on the number of available end depots. 
\end{itemize}

\noindent Although using our formulation in a commercial solver still cannot match the performance of a dedicated implementation of the DP approach of \citet{Ratliff:1983} in a higher programming language, the former approach has the following advantages: 
\begin{itemize}
\item The formulation can be easily be implemented and used by anyone familiar with a mathematical programming solver. No knowledge of a higher programming language is required, and no experience in algorithmic programming to realize a DP is necessary. This point is certainly relevant in practice, where algorithmic programming skills are generally far rarer than at universities and other scientific organizations. 
\item The formulation is extendable to handle three important settings in modern e-commerce warehouses---scattered storage, decoupling of picker and cart, and multiple end depots---and seems likely to be able to incorporate other real-world-inspired constraints.  
\item The formulation can be used in approaches for integrated problems, in which the higher-level decision depends on the outcome of the SPRP, e.g., order batching \citep{Gademann:2005, Valle:2017} or storage assignment \citep{Petersen:1999}. For example, the integrated order batching and picker routing problem could be solved by i)~column generation, where our model can be modified to solve the pricing subproblem, i.e., the orders are associated with current prices, and the picker can only pick orders such that the number of collected items does not exceed the maximum batch size, or ii)~by a compact formulation that extends our model by an index for each batch (up to the maximum number of batches), a limited capacity for each batch, and a set covering constraint. The integrated storage assignment and picker routing problem could be studied in a scattered-storage setting using our model.
\end{itemize}
This paper is organized as follows. We introduce our compact formulation for the standard SPRP in Section~\ref{sec:model}. The following sections present the extensions of the model to the setting with scattered storage (Section~\ref{sec:mixshelves}), decoupling of picker and cart (Section~\ref{sec:cart}), and multiple end depots (Section~\ref{sec:openDepot}). Section~\ref{sec:results} presents the numerical studies to investigate the performance of our formulation and the benefits of the considered extensions. Section~\ref{sec:conclusion} concludes the paper.  

\section{Mathematical Formulation of the Standard SPRP}
\label{sec:model}
To solve an instance of the standard SPRP, the warehouse can be restricted to its relevant part, i.e., all aisles that lie to the left of both the depot and the leftmost aisle in which a SKU needs to be picked, and, analogously, all aisles that lie to the right of both the depot and the rightmost aisle in which a SKU needs to be picked, can be removed. The resulting part of the warehouse is represented as a set $\setOfAisles=\{ \firstAisle, \ldots,  \lastAisle \}$ indexing $\numberOfAisles$ aisles numbered from left to right. Each aisle $\aisle \in \setOfAisles$ has $n$ available picking positions, numbered from top to bottom, and is associated with a set of required picking positions $\setOfArticlesInAisle{\aisle} \subseteq \{0,\ldots, n-1\}$ that the picker needs to visit to complete the pick list. The depot can be located at the entries to the picking aisles in the top or bottom cross aisle. The picking aisle above\,/\,below which the depot is located is denoted as aisle $\depotAisle$. 
The parameter $\depotTop{}$ is set to 1 if the depot is located in the top cross aisle and to zero otherwise.

\begin{figure}[htbp]
	\centering
  \includegraphics[width=0.8\textwidth]{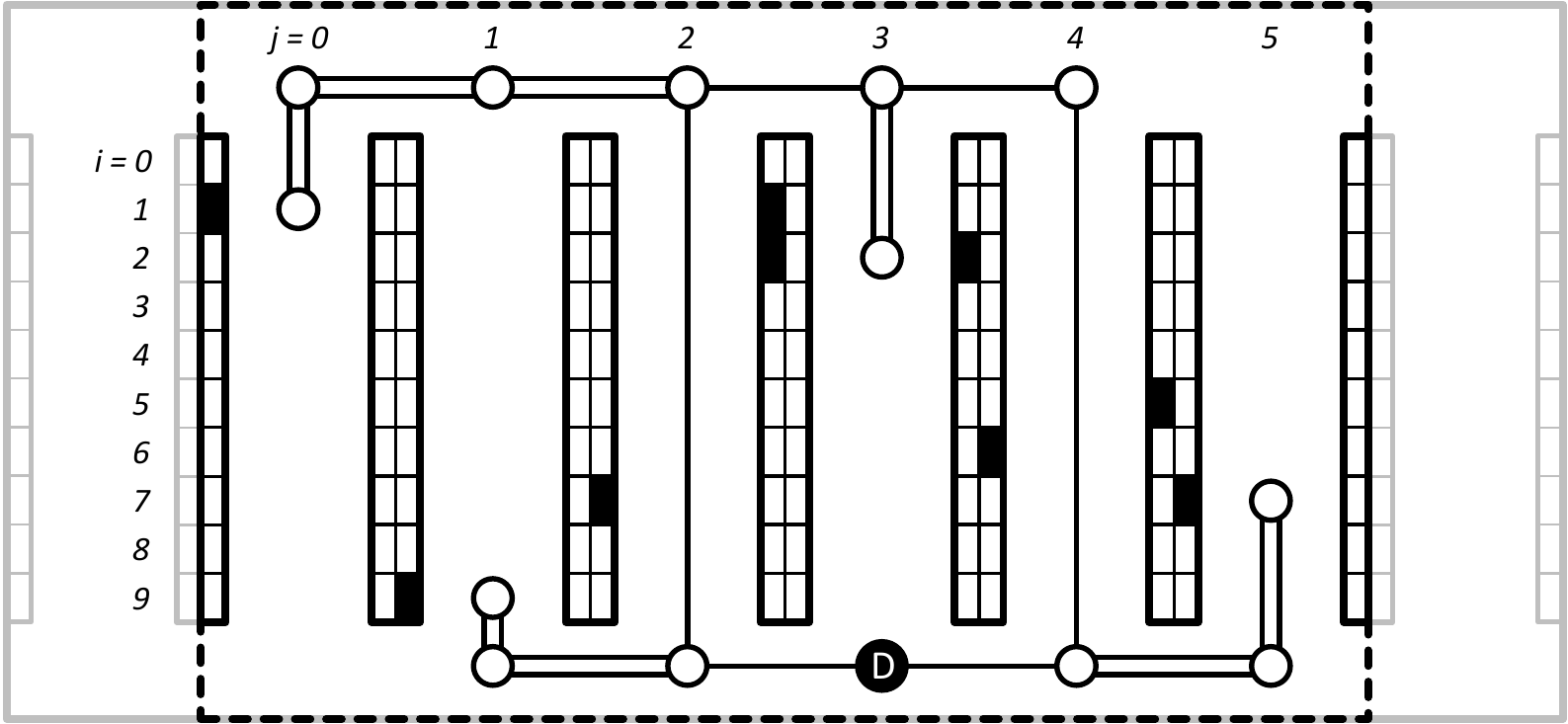}
	\caption{Optimal solution of an example instance of the standard SPRP.}
	\label{fig:warehouse}
\end{figure}

The example in Figure~\ref{fig:warehouse} illustrates the introduced concepts: There are eight picking aisles with $n=10$ available picking positions per aisle. SKUs need to be picked from nine required picking positions that are marked black. We only have to consider the $\numberOfAisles = 6$ aisles containing required picking positions, i.e.,  $\setOfAisles=\{0, \ldots 5\}$. The required picking positions in the aisles are given by $\setOfArticlesInAisle{0}=\{1\}$, $\setOfArticlesInAisle{1}=\{9\}$, $\setOfArticlesInAisle{2}=\{2,3,7\}$, $\setOfArticlesInAisle{3}=\{3\}$, $\setOfArticlesInAisle{4}=\{5,6\}$, and $\setOfArticlesInAisle{5}=\{7\}$. The depot is located at the bottom of aisle 3, i.e., $\depotAisle=3$ and $\depotTop{} = 0$. 

As described by \cite{Ratliff:1983}, there exist four feasible configurations to connect aisle $\aisle$ and aisle $\nextAisle$ using the cross aisles located at the top and bottom of the warehouse. 
We represent these configurations by means of the following binary decision variables:
\begin{itemize}
\item $\xTwoTop{\aisle}$ equals 1 if the top cross aisle is traversed twice (back and forth), 0 otherwise,
\item $\xTwoBottom{\aisle}$ equals 1 if the bottom cross aisle is traversed twice (back and forth), 0 otherwise,
\item $\xTopBottom{\aisle}$ equals 1 if the bottom and the top cross aisle are both traversed once, 0 otherwise, and
\item $\xTwoTopBottom{\aisle}$ equals 1 if both the top and bottom cross aisle are both traversed twice, 0 otherwise.
\end{itemize}
With regard to the traversal of picking aisles, the binary decision variables $\xUp{\aisle}$ are used to indicate that aisle $\aisle$ is completely traversed once (in arbitrary direction). If the costs for traversing the picking aisles are non-uniform, it may be beneficial to traverse an aisle $\aisle$ twice: in this case, $\xTwoUp{\aisle}$ equals one. For uniform traversal costs, there always exists an optimal solution in which no aisle is traversed twice \citep[see][]{Ratliff:1983}. Furthermore, the binary decision variables $\xPickTop{\aisle}{\article}$ ($\xPickBottom{\aisle}{\article}$) define that picking position $\article$---and all picking positions that are located between the top (bottom) and picking position $\article$---are accessed via a \vertical branch-and-pick tour from the top (bottom) of aisle $\aisle$, i.e., the picking aisle is entered and left from the same cross aisle. For each of the described decision variables, we precompute cost coefficients $\cost{}$ that correspond to the additional travel cost of the picker if the respective decision variable equals 1, e.g., $\xTwoTopBottom{\aisle}$ has a coefficient $\costTwoTopBottom{\aisle}$ that corresponds to four times the travel cost between aisle $\aisle$ and $\nextAisle$. Figure~\ref{fig:warehouse} illustrates the meaning of the decision variables: here, $\xPickTop{0}{1}, \xTwoTop{0}, \xPickBottom{1}{9}, \xTwoTopBottom{1}, \xUp{2}, \xTopBottom{2},\xPickTop{3}{2}, \xTopBottom{3}, \xUp{4},\xTwoBottom{4}, \xPickBottom{5}{7}$ are equal to $1$. 

According to \citet{Ratliff:1983}, the generation of isolated subtours is prevented if the degree of the connections at the top and bottom of each picking aisle is of even, and the picking tour is connected. Using the observation that an even degree divided by two is an integer, we introduce for each picking aisle $\aisle$ an integer variable $\degreeEvenTop{\aisle}$ ($\degreeEvenBottom{\aisle}$), whose value corresponds to the degree of the connections at the top (at the bottom) of aisle $\aisle$ divided by two. For example, $\degreeEvenTop{2} = 2$ and $\degreeEvenBottom{3} = 1$  in Figure~\ref{fig:warehouse}. To guarantee that the picking tour is connected, we introduce an additional binary variable $\xComponent{\aisle}$ for each picking aisle $\aisle$, which is equal to 0 if the picking tour from the leftmost relevant aisle in the warehouse up to aisle $\aisle$ is connected, and equal to 1 if this picking tour consists of two components. Note that two components emerge whenever (i)~we have configuration $\xTwoTopBottom{\aisle}$ for the leftmost aisle, or (ii)~we switch from configuration $\xTwoTop{\previousAisle}$ or $\xTwoBottom{\previousAisle}$ to configuration $\xTwoTopBottom{\aisle}$ without connecting the top and bottom cross aisle.

Using the above definitions, we can formulate the standard SPRP. To present a concise and comprehensible model (by avoiding the repetition of basically identical constraints for different index sets), we take the following two liberties with regard to notation: (i)~we sometimes use conditional statements for defining the relevant index set or to define the validity of a certain set of constraints, and (ii)~ we use the notation $\underset{\mathclap{\mathit{range}}}{[}\ldots]$ to define that a certain part of an expression is only relevant for a given range of the index (and otherwise disappears). 

\begin{align}
\text{min~}&\sum_{\aisle \in \setOfAisles} \costTwoBottom{\aisle} \xTwoBottom{\aisle} + \costTwoTop{\aisle} \xTwoTop{\aisle} + \costTopBottom{\aisle} \xTopBottom{\aisle}  + \costTwoTopBottom{\aisle} \xTwoTopBottom{\aisle} +  \costUp{\aisle} \xUp{\aisle} + \costTwoUp{\aisle} \xTwoUp{\aisle}+  
\sum_{\aisle \in \setOfAisles}\sum_{\article \in \setOfArticlesInAisle{\aisle}} \left(\costPickBottom{\aisle}{\article} \xPickBottom{\aisle}{\article} +\costPickTop{\aisle}{\article} \xPickTop{\aisle}{\article}\right)
&~~~~~~~~~~~~~~~~~~~~~~~~\label{F0}
\end{align}
{\begin{align}
%
%
\text{s.t.~}&\xTwoBottom{\aisle} +  \xTwoTop{\aisle} +  \xTopBottom{\aisle}  + \xTwoTopBottom{\aisle}  = 1 & \aisle \in \setOfAisles\setminus \{\lastAisleVis\} \label{F1}\\
 &\xUp{\aisle} +  \xTwoUp{\aisle} +  \sum_{\article' \in \setOfArticlesInAisle{\aisle} :\article'\geq\article} \xPickTop{\aisle}{\article'}  + \sum_{\article' \in \setOfArticlesInAisle{\aisle} :\article'\leq\article}\xPickBottom{\aisle}{\article'} \geq 1 & \aisle \in \setOfAisles, \article \in \setOfArticlesInAisle{\aisle} \label{F2}\\
&\underset{\mathclap{j > \firstAisle}}{[}\xTwoBottom{\previousAisle} +   \xTopBottom{\previousAisle}  + \xTwoTopBottom{\previousAisle}]+ \xTwoBottom{\aisle} + \xTopBottom{\aisle}  + \xTwoTopBottom{\aisle}   \geq \xPickBottom{\aisle}{\article} & \begin{aligned}[c] \mathit{if}(\depotTop{} = 1)\,\{\aisle \in \setOfAisles\}\,\\\mathit{else}\,\{\aisle \in \setOfAisles \setminus \{\depotAisle\} \}, \article \in \setOfArticlesInAisle{\aisle} \end{aligned} \label{F3}\\[15pt]
&  \underset{\mathclap{j > \firstAisle}}{[}
\xTwoTop{\previousAisle} +  \xTopBottom{\previousAisle}  + \xTwoTopBottom{\previousAisle}] +  \xTwoTop{\aisle} +  \xTopBottom{\aisle}  + \xTwoTopBottom{\aisle}  \geq \xPickTop{\aisle}{\article} & \begin{aligned}[c] \mathit{if}(\depotTop{} = 0)\,\{\aisle \in \setOfAisles\}\,\\\mathit{else}\,\{\aisle \in \setOfAisles \setminus \{\depotAisle\} \}, \article \in \setOfArticlesInAisle{\aisle} \end{aligned}  \label{F4}\\[15pt]
&\xTwoTop{\previousAisle} + \xTwoBottom{\aisle} \leq \xTwoUp{\aisle}+1 & \aisle \in \setOfAisles \setminus \{\firstAisle\}\label{F5}\\
&\xTwoBottom{\previousAisle} + \xTwoTop{\aisle} \leq \xTwoUp{\aisle}+1 & \aisle \in \setOfAisles \setminus \{\firstAisle\}\label{F6}\\
&2 \xTwoUp{\depotAisle} + \xUp{\depotAisle} +\underset{\mathclap{\depotAisle > \firstAisle}}{[}\xTwoTop{\depotAisle-1} +\xTwoTopBottom{\depotAisle-1}  ] + \xTwoTop{\depotAisle} +\xTwoTopBottom{\depotAisle} \geq  \underset{\mathclap{\depotAisle > \firstAisle}}{[}\xTwoBottom{\depotAisle-1}] + \xTwoBottom{\depotAisle}& \mathit{if}(\depotTop{} = 1) \label{F7}\\
&2 \xTwoUp{\depotAisle} + \xUp{\depotAisle} +\underset{\mathclap{\depotAisle > \firstAisle}}{[}\xTwoBottom{\depotAisle-1} + \xTwoTopBottom{\depotAisle-1}] + \xTwoBottom{\depotAisle} +\xTwoTopBottom{\depotAisle} \geq  \underset{\mathclap{\depotAisle > \firstAisle}}{[}\xTwoTop{\depotAisle-1}] + \xTwoTop{\depotAisle}& \mathit{if}(\depotTop{}= 0)\label{F8}\\
%
%
&\underset{\mathclap{j > \firstAisle}}{[}\xTopBottom{\previousAisle} +
2 \xTwoTopBottom{\previousAisle} + 
2 \xTwoTop{\previousAisle} ] +  \xTopBottom{\aisle} + 
2 \xTwoTopBottom{\aisle}  +
2 \xTwoTop{\aisle} +
\xUp{\aisle}+
2 \xTwoUp{\aisle}
= 2 \degreeEvenTop{\aisle} & \aisle \in \setOfAisles \label{E1}\\
&\underset{\mathclap{j > \firstAisle}}{[}\xTopBottom{\previousAisle} +
2 \xTwoTopBottom{\previousAisle} + 
2 \xTwoBottom{\previousAisle} ] +  \xTopBottom{\aisle} + 
2 \xTwoTopBottom{\aisle}  +
2 \xTwoBottom{\aisle} +
\xUp{\aisle}+
2 \xTwoUp{\aisle}
= 2 \degreeEvenBottom{\aisle} & \aisle \in \setOfAisles \label{E2}\\
%
%
&\xTwoTopBottom{\aisle} + \xTwoBottom{\previousAisle} + \xTwoTop{\previousAisle} - \xTwoUp{\aisle}  \leq \xComponent{\aisle} +1 & \aisle \in \setOfAisles \setminus \{\firstAisle\}\label{T1}\\
&\xTwoTopBottom{\aisle}+\underset{\mathclap{j > \firstAisle}}{[}-\xTwoTopBottom{\previousAisle}-\xTwoTop{\previousAisle} -\xTwoBottom{\previousAisle}] - \xTwoUp{\aisle} - \xUp{\aisle}  \leq \xComponent{\aisle} & \aisle \in \setOfAisles \label{T3}\\
&\xComponent{\previousAisle}  - \xUp{\aisle}- \xTwoUp{\aisle} \leq \xComponent{\aisle}  & \aisle \in \setOfAisles \setminus \{\firstAisle\}\label{T4}\\
&\xComponent{\aisle} \leq \xTwoTopBottom{\aisle}  & \aisle \in \setOfAisles \label{T5}\\
%
%
&\xTopBottom{\aisle}, \xTwoTop{\aisle}, \xTwoBottom{\aisle}, \xTwoTopBottom{\aisle}, \xComponent{\aisle} \in \{0,1\} & \aisle \in \setOfAisles \setminus \{\lastAisle\} \label{B1}\\
&\xUp{\aisle}, \xTwoUp{\aisle} \in \{0,1\} & \aisle \in \setOfAisles \label{B2}\\
& \xPickTop{\aisle}{\article}, \xPickBottom{\aisle}{\article}\in \{0,1\} & \aisle \in \setOfAisles, \article \in \setOfArticlesInAisle{\aisle} \label{B3}\\
& \degreeEvenTop{\aisle},\degreeEvenBottom{\aisle} \in \mathcal{N}_{0} & \aisle \in \setOfAisles \label{B4}\\
& \xTopBottom{\lastAisle}, \xTwoTop{\lastAisle}, \xTwoBottom{\lastAisle}, \xTwoTopBottom{\lastAisle}, \xComponent{\lastAisle} = 0 &  \label{B5}
\end{align}}
The objective~\eqref{F0} is to minimize the total costs of the picker tour. Constraints~\eqref{F1} guarantee that the relevant part of the warehouse is visited by the picker using one of the four cross aisle configurations. Constraints~\eqref{F2} ensure that the picker visits all required picking positions. Constraints~\eqref{F3} and \eqref{F4} guarantee that a \vertical branch-and-pick tour from the bottom (top) cross aisle into aisle $\aisle$ can only take place if $\aisle$ is connected to the previous or successive aisle with a configuration that uses the bottom (top) cross aisle. The squared brackets exclude the terms involving the preceding aisle when the constraints for the first picking aisle are determined. Constraints~\eqref{F5} and \eqref{F6} guarantee that switches between top and bottom cross aisle are connected in a feasible manner. Constraints~\eqref{F7} and \eqref{F8} ensure that the depot is included in the tour. For example, if the depot is located at the top $(\depotTop{} = 1)$ and the aisle containing the depot is connected via $\xTwoBottom{\depotAisle} =1$ and $\xTwoBottom{\depotAisle-1} =1$, the depot must be included in the tour by setting $\xTwoUp{\depotAisle} =1$. Instead, if $\xTwoBottom{\depotAisle} =1$ and $\xTwoBottom{\depotAisle-1} =0$, the depot must be included by setting $\xTwoUp{\depotAisle} =1$, $\xTwoTopBottom{\depotAisle-1} =1$, or $\xTwoTop{\depotAisle-1}=1$. Constraints~\eqref{E1} and \eqref{E2} establish that the degrees of all connections at the top and also at the bottom of each picking aisle must be even, i.e, every position must be left as often as it is entered. 
Constraints~\eqref{T1} set the number of components to two (i.e., $\xComponent{\aisle}=1$) if there is a transition from configurations $\xTwoBottom{\aisle-1} =1$ or $\xTwoTop{\aisle-1} =1$ to $\xTwoTopBottom{\aisle} =1$ without directly connecting top and bottom by $\xTwoUp{\aisle} =1$. Constraints~\eqref{T3} set the number of components to two, if top and bottom are not connected by a traversal of the picking aisle, and the part of the warehouse to the left is not visited. Constraints~\eqref{T4} propagate the number of components. Constraints~(\ref{T5}) ensure that configuration $\xTwoTopBottom{\aisle}$ is used as long as there are two components.
Finally, constraints~\eqref{B1}--\eqref{B5} define the decision variables.

Note that the model could be further improved by substituting for every aisle $\aisle$ the variables $\xPickBottom{\aisle}{\article}, \xPickTop{\aisle}{\article}, \article \in \setOfArticlesInAisle{\aisle}$ with three new variables 
 $\xPickBottom{}{\aisle}, \xPickTop{}{\aisle},\xPickTopBottom{\aisle}$ that represent the three \vertical branch-and-pick tours that can alternatively be part of an optimal tour, i.e, from the bottom cross aisle to the topmost requested SKU, from the top cross aisle to the bottommost requested SKU, and from top cross aisle and bottom cross aisle to the two neighboring requested SKUs with the largest distance between them. We refrained from implementing this improvement to keep a more general formulation as a basis for the extensions presented in the following sections.

\section{The Single-Block SPRP with Scattered Storage}
\label{sec:mixshelves}
\begin{figure}[htbp]
	\centering
  \includegraphics[width=0.8\textwidth]{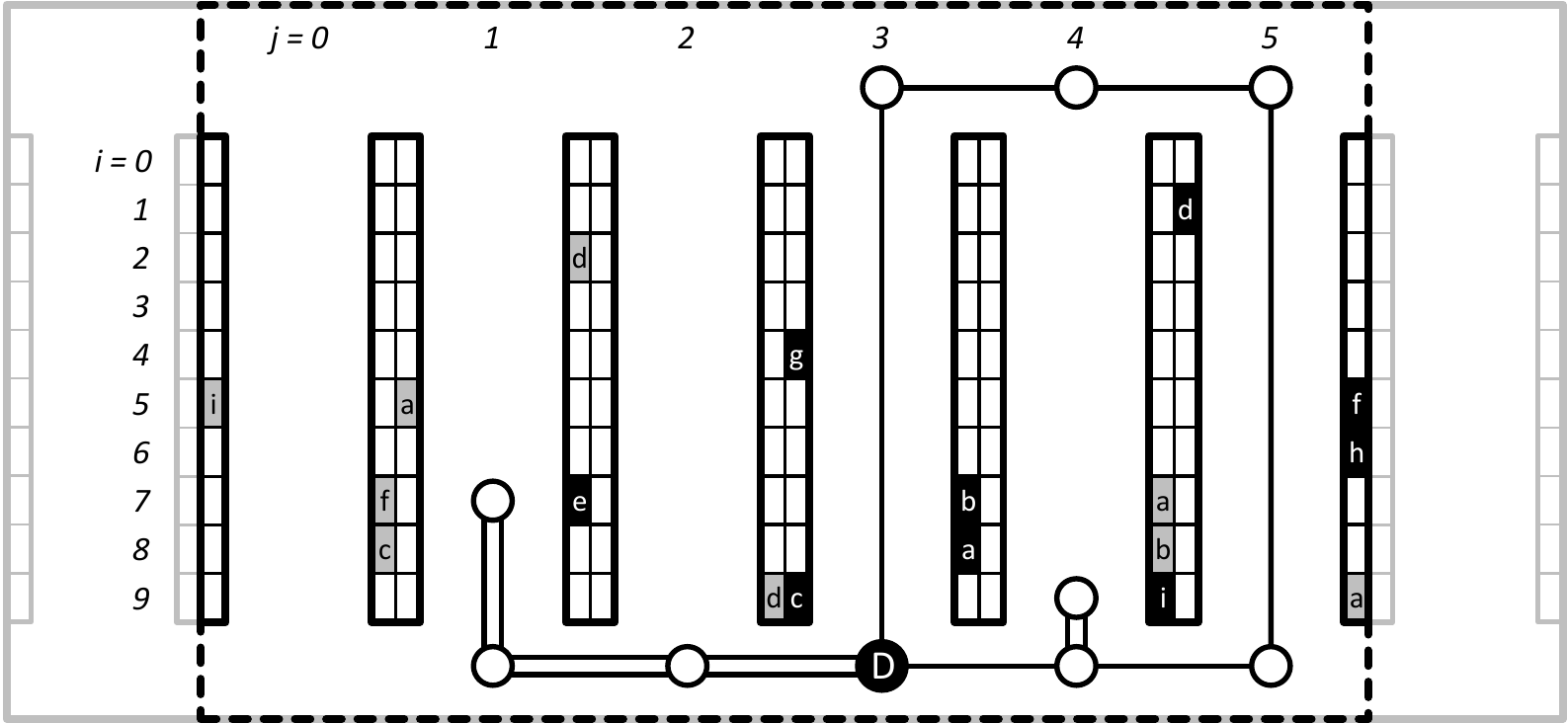}
	\caption{Optimal solution of an example instance of the single-block SPRP with scattered storage. Picking positions from which a requested SKU is picked are marked in black.}
	\label{fig:warehouse2}
\end{figure}

We extend formulation~\eqref{F0}--\eqref{B5} to the single-block SPRP with scattered storage, i.e., now, any SKU can be available from multiple picking positions. Figure~\ref{fig:warehouse2} shows the optimal solution of an example instance of this problem variant. We assume that multiple items of each individual SKU may be contained in the pick list and that the supply of items of a SKU available at a given picking position is limited. Set $\setOfArticles$ contains all SKUs that need to be picked, and  $\demand{\myArticle}$ denotes the number of items of SKU $\myArticle \in \setOfArticles$ that are requested. Set $\setOfCellsInAisleForArticle{\aisle}{\myArticle}$ contains all picking positions from which SKU $\myArticle$ is available in aisle $\aisle$, and $\capacityTwo{\aisle}{\article}{\myArticle}$ denotes the number of items of SKU $\myArticle$ that are available in aisle $\aisle$ at position $\article$. Set $\setOfArticlesInAisle{\aisle}$ is redefined to contain all picking positions in aisle $\aisle$ from which SKUs present in the pick list are available: several positions storing the same SKU may be contained, and not all positions have to be visited. To indicate whether picking position $\article$ in aisle ${\aisle}$ is visited, we introduce additional binary variables $\xPick{\aisle}{\article}$. In the example in Figure~\ref{fig:warehouse2}, we assume $\setOfArticles = \{a,b,c,d,e,f,g,h,i\}$, $\capacityTwo{\aisle}{\article}{\myArticle} = 1,\myArticle \in \setOfArticles, \aisle \in \setOfAisles, \article \in \setOfArticlesInAisle{\aisle}$, and $b_h = 1,\myArticle \in \setOfArticles$. The picking tour is given by $\xPickBottom{1}{7}, \xTwoBottom{1}, \xTwoBottom{2}, \xUp{3}, \xTopBottom{3}, \xPickBottom{4}{9}, \xTopBottom{4},\xUp{5}$ equal to $1$, and, e.g., $\xPick{3}{8} = 1$ indicates that the requested item of SKU 'a' is picked in aisle 3 at position 8. 

To model the decision where to pick the requested SKUs,  we replace constraints~\eqref{F2} by constraints~\eqref{F2c}--\eqref{F2d}:
\begin{align} 
\tiny
&\sum_{\aisle \in \setOfAisles} \sum_{\article \in \setOfCellsInAisleForArticle{\aisle}{\myArticle}}  \capacityTwo{\aisle}{\article}{\myArticle} \xPick{\aisle}{\article} \geq \demand{\myArticle} & \myArticle \in  \setOfArticles \label{F2c} \\
&\xUp{\aisle} +  \xTwoUp{\aisle} +  \sum_{\article' \in \setOfArticlesInAisle{\aisle} :\article'\geq\article} \xPickTop{\aisle}{\article'}  + \sum_{\article' \in \setOfArticlesInAisle{\aisle} :\article'\leq\article}\xPickBottom{\aisle}{\article'} \geq \xPick{\aisle}{\article}\ & \aisle \in \setOfAisles, \article \in \setOfArticlesInAisle{\aisle} \label{F2b} \\
&\xPick{\aisle}{\article} \in \{0,1\} & \aisle \in \setOfAisles, \article \in \setOfArticlesInAisle{\aisle} \label{F2d}
\end{align}
Constraints~\eqref{F2c} ensure that the requested number of items of each SKU is picked from the picking positions at which the SKU is available. Constraints~\eqref{F2b} guarantee that the selected positions are visited by the picking tour.  

Because not all picking positions storing requested SKUs have to be visited in the case of scattered storage, it is not possible to define the relevant part of the warehouse by means of {these} picking positions and the location of the depot like in the standard SPRP. Instead, we require  additional binary variables $\xLast{\aisle}$ that indicate whether aisle $\aisle$ is reached by the picker or not, i.e., in Figure~\ref{fig:warehouse2}, $\xLast{1}, ..., \xLast{5}$ are equal to $1$. We replace constraints~\eqref{F1} by the following constraints: 
\begin{align} 
\tiny
& \xLast{\aisle} \geq \xPick{\aisle}{\article} & \aisle \in \setOfAisles, \article \in \setOfArticlesInAisle{\aisle} \label{F1e} \\
&\xLast{l} = 1 &  \label{F1fb} \\
&\xTwoBottom{\aisle} +  \xTwoTop{\aisle} +  \xTopBottom{\aisle}  + \xTwoTopBottom{\aisle}  = \xLast{\aisle+1} & \aisle \in \setOfAisles\setminus \{\lastAisleVis\} : j \geq l \label{F1b} \\
&\xTwoBottom{\aisle} +  \xTwoTop{\aisle} +  \xTopBottom{\aisle}  + \xTwoTopBottom{\aisle}  = \xLast{\aisle} & \aisle \in \setOfAisles : j < l \label{F1bb} \\
& \xLast{\aisle} \geq \xLast{\aisle+1} & \aisle \in \setOfAisles\setminus \{\lastAisleVis\} : j \geq l \label{F1d} \\
& \xLast{\aisle} \leq \xLast{\aisle+1} & \aisle \in \setOfAisles: j < l \label{F1db} \\
&\xLast{\aisle} \in \{0,1\} & \aisle \in \setOfAisles \label{F1f}
\end{align}
Constraints~\eqref{F1e} and \eqref{F1fb} guarantee that all aisles containing the selected picking positions of the requested SKUs and the aisle containing the depot are reached. Constraints~\eqref{F1b} define the configurations to connect aisles that allow reaching a certain aisle located to the right of the depot, and \eqref{F1bb} does the same for the aisles to the left of the depot. Constraints~\eqref{F1d} and \eqref{F1db} describe how to propagate the $\xLast{}$ variables.

\section{The Single-Block SPRP with Decoupling}
\label{sec:cart}
In this section, we extend our model to cover the possibility of the picker to park the cart and continue the tour for a certain period without the cart, then returning to the cart. Figure~\ref{fig:warehouse3} depicts an example of an optimal solution for the resulting single-block SPRP with decoupling. We make the following modeling assumptions:

\begin{figure}[htbp]
	\centering
  \includegraphics[width=0.8\textwidth]{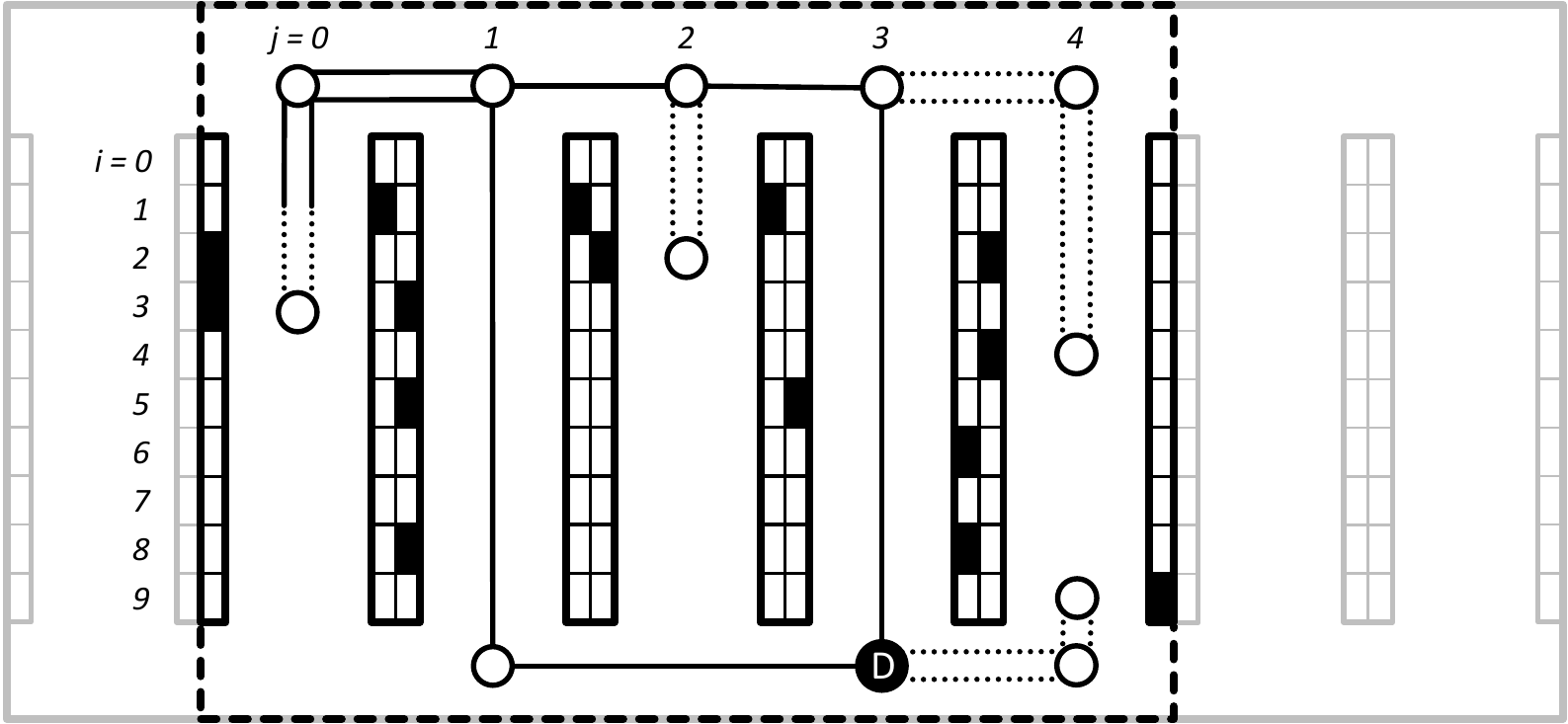}
	\caption{Optimal solution of an example instance of the single-block SPRP with decoupling. Only one SKU is stored at each picking position, and only one item is requested of each SKU. The picker capacity is two items, and the speed of the picker without cart is twice the speed with cart. The tours that the picker does without cart are indicated by dotted edges.}
	\label{fig:warehouse3}
\end{figure}

\begin{itemize}
\item The speed of the picker without cart differs from the speed when pushing the cart. The cost coefficients $\cost{}$ now represent the travel times of the picker when pushing the cart, the coefficients $\cost{}^p$ the times of the picker when traveling alone. 
\item Like in the previous models, we assume that the pick list is generated such that the capacity of the picking cart is sufficient to carry all items. However, the carrying capacity of the picker alone is limited to $C$ items. We return to a warehouse with dedicated storage, i.e., each SKU is only stored at one picking position, and $\demandTwo{\aisle}{\article}{\myArticle}$ denotes the number of requested items of SKU $\myArticle$ that have to be picked from aisle $\aisle$ at position $\article$.
\item If a picking aisle is completely traversed, the picker pushes the cart, and no decoupling takes place.  
\item In \horizontal branch-and-pick tours, the picker alone travels along a cross aisle and may visit one or more picking aisles to retrieve SKUs. In Figure~\ref{fig:warehouse3}, one \horizontal branch-and-pick tour starts at the top of aisle 3, another one at the bottom. We assume that after parking the cart, at most one \vertical branch-and-pick tour (see Figure~\ref{fig:warehouse3}, aisles 0 and 2 for examples of \vertical branch-and-pick tours) and at most two \horizontal branch-and-pick tours---one to the left and one to the right---are possible. Note that this assumption is not restrictive if the picker without cart is not more than twice as fast as the picker pushing the cart. Lifting the assumption is not possible with the presented modeling approach because it would entail that a section of an aisle is traversed more than twice. For the same reason, we assume that branch-and-pick tours (\vertical as well as \horizontal) starting from different parking positions cannot overlap.  

Parking the cart directly at the entry or within a picking aisle and doing a \vertical branch-and-pick tour in that aisle does not explicitly have to be modeled: Under our assumption that only one \vertical branch-and-pick tour for each parking position is allowed, it is beneficial to only push the cart as far into the picking aisle as is needed so that the capacity of the picker is sufficient to visit the remaining required picking positions in the aisle (see Figure~\ref{fig:warehouse3}, aisle 0). Because the decision $\xPickTop{\aisle}{\article}=1$ ($\xPickBottom{\aisle}{\article}=1$)  implicates that all positions located above (below) position $\article$ in aisle $\aisle$ are picked in the branch-and-pick tour, the parking position of the cart and consequently the respective cost coefficients $\costPickTop{\aisle}{\article}$ and $\costPickBottom{\aisle}{\article}$ can be precomputed. However, decoupling and the parking position of the cart must explicitly be modeled for \horizontal branch-and-pick tours. In this case, possible parking positions are located in the cross aisles at the entries to the picking aisles.
\end{itemize}
To model the path of the picker when traveling without cart, we introduce additional binary variables: Variables $\xPickerTopRight{\aisle}$ indicate that the picker traverses the top cross aisle from the entry of picking aisle $\aisle$ to the entry of $\aisle+1$, and the cart is parked somewhere to his left, i.e., the picker is traveling to the right. Variables $\xPickerBottomRight{\aisle}$ are defined analogously for the bottom cross aisle. Variables $\xPickerBottomLeft{\aisle}$ and $\xPickerTopLeft{\aisle}$ indicate that the picker is traveling from the entry of aisle $\aisle+1$ to the entry of $\aisle$, and the cart is parked to his right. Variables $\xPickTop{\aisle}{\article}^p$ ($\xPickBottom{\aisle}{\article}^p$) indicate that picking aisle $\aisle$ is entered from the top (bottom) without cart, and all required picking positions down (up) to position $\article$ are visited. The variables are only defined for those picking positions $i$ for which the picker capacity is sufficient to carry the requested number of items of all SKUs that are stored in the picking positions passed by the picker. In the example in Figure~\ref{fig:warehouse3}, variables $\xPickTop{0}{3}, \xTwoTop{0}, \xUp{1}, \xTopBottom{1}, \xPickTop{2}{2}, \xTopBottom{2}, \xUp{3}$ are equal to $1$ and define the picking tour with cart, and variables $\xPickerTopRight{3}, \xPickTop{4}{4}^p, \xPickerBottomRight{3}, \xPickBottom{4}{9}^p$ equal $1$ and describe the two \horizontal branch-and-pick tours starting from aisle 3. 

We modify formulation~\eqref{F0}--\eqref{B5} as follows: First, the objective function~\eqref{F0} is changed to minimize the total travel time, i.e., the sum of the time the picker travels alone and the time that the picker travels with cart:
\begin{align}
\tiny
& \begin{aligned}
\text{min~}&\sum_{\aisle \in \setOfAisles} \costTwoBottom{\aisle} \xTwoBottom{\aisle} + \costTwoBottom{\aisle}^p (\xPickerBottomLeft{\aisle} + \xPickerBottomRight{\aisle}) + \costTwoTop{\aisle} \xTwoTop{\aisle} +  \costTwoTop{\aisle}^p (\xPickerTopLeft{\aisle} + \xPickerTopRight{\aisle}) + \costTopBottom{\aisle} \xTopBottom{\aisle}  + \costTwoTopBottom{\aisle} \xTwoTopBottom{\aisle} +  \costUp{\aisle} \xUp{\aisle} + \costTwoUp{\aisle} \xTwoUp{\aisle}+  \\
&\sum_{\aisle \in \setOfAisles}\sum_{\article \in \setOfArticlesInAisle{\aisle}} \left(\costPickBottom{\aisle}{\article} \xPickBottom{\aisle}{\article} +\costPickTop{\aisle}{\article} \xPickTop{\aisle}{\article}\right) +  \left(\costPickBottom{\aisle}{\article}^p \xPickBottom{\aisle}{\article}^p +\costPickTop{\aisle}{\article}^p \xPickTop{\aisle}{\article}^p\right)
 \end{aligned} & \label{H4}
 \end{align}

\noindent We replace constraints~\eqref{F2} with constraints~\eqref{F2Pick} to take picking without cart into account:
\begin{align} 
\tiny
 &\xUp{\aisle} +  \xTwoUp{\aisle} +  \sum_{\article' \in \setOfArticlesInAisle{\aisle} :\article'\geq\article} (\xPickTop{\aisle}{\article'} + \xPickTop{\aisle}{\article'}^p)  + \sum_{\article' \in \setOfArticlesInAisle{\aisle} :\article'\leq\article} (\xPickBottom{\aisle}{\article'} + \xPickBottom{\aisle}{\article'}^p) \geq 1 & \aisle \in \setOfAisles, \article \in \setOfArticlesInAisle{\aisle} \label{F2Pick}
\end{align}
To determine the part of the warehouse in which the cart is used, we replace constraints~\eqref{F1} with constraints~\eqref{F1fb}--\eqref{F1f}, and we add the following constraints:
\begin{align} 
\tiny
&\xLast{\aisle} \geq \xTwoUp{\aisle} &  \aisle \in \setOfAisles \label{F1z}
\end{align} 

\noindent To model feasible \horizontal branch-and-pick tours of the picker without cart, we add: 
\begin{align} 
\tiny
&\xPickerBottomRight{\aisle} + \xPickerBottomLeft{\aisle} + \xTwoBottom{\aisle} + \xTwoTopBottom{\aisle} + \xTopBottom{\aisle} \leq  1 & \aisle \in \setOfAisles\setminus \{\lastAisleVis\} \label{G1} \\
&\xPickerTopRight{\aisle} + \xPickerTopLeft{\aisle} +  \xTwoTop{\aisle} + \xTwoTopBottom{\aisle} + \xTopBottom{\aisle} \leq  1 & \aisle \in \setOfAisles\setminus \{\lastAisleVis\} \label{G2} \\
&\underset{\mathclap{j > \firstAisle}}{[}\xPickerTopRight{\previousAisle}  +  \xTwoTop{\previousAisle} + \xTwoTopBottom{\previousAisle}] + \xUp{\aisle} +\xTwoUp{\aisle} \geq  \xPickerTopRight{\aisle} & \aisle \in \setOfAisles\setminus \{\lastAisleVis\}\label{G3} \\
&\underset{\mathclap{j > \firstAisle}}{[}\xPickerBottomRight{\previousAisle} +  \xTwoBottom{\previousAisle} + \xTwoTopBottom{\previousAisle}] + \xUp{\aisle} +\xTwoUp{\aisle} \geq  \xPickerBottomRight{\aisle} & \aisle \in \setOfAisles\setminus \{\lastAisleVis\}\label{G4} \\
&\xPickerTopLeft{\aisle+1}  +  \xTwoTop{\aisle+1} + \xTwoTopBottom{\aisle+1} + \xUp{\aisle+1} +\xTwoUp{\aisle+1} \geq  \xPickerTopLeft{\aisle} & \aisle \in \setOfAisles\setminus \{\lastAisleVis\} \label{G5} \\
&\xPickerBottomLeft{\aisle+1}  +  \xTwoBottom{\aisle+1} + \xTwoTopBottom{\aisle+1} + \xUp{\aisle+1} +\xTwoUp{\aisle+1} \geq  \xPickerBottomLeft{\aisle} & \aisle \in \setOfAisles\setminus \{\lastAisleVis\} \label{G6} \\
&\underset{\mathclap{j > \firstAisle}}{[}\xPickerBottomRight{\previousAisle}] +   \xPickerBottomLeft{\aisle} \geq \xPickBottom{\aisle}{\article}^p & \aisle \in \setOfAisles, \article \in \setOfArticlesInAisle{\aisle} \label{G7}\\
&\underset{\mathclap{j > \firstAisle}}{[}\xPickerTopRight{\previousAisle}] +   \xPickerTopLeft{\aisle} \geq \xPickTop{\aisle}{\article}^p & \aisle \in \setOfAisles, \article \in \setOfArticlesInAisle{\aisle} \label{G8}\\
& \xPickerBottomRight{\aisle},\xPickerBottomLeft{\aisle},\xPickerTopRight{\aisle}, \xPickerTopLeft{\aisle} \in \{0,1\} & \aisle \in \setOfAisles \setminus \{\lastAisleVis\} \label{G9}\\
& \xPickerBottomRight{\lastAisleVis},\xPickerBottomLeft{\lastAisleVis},\xPickerTopRight{\lastAisleVis}, \xPickerTopLeft{\lastAisleVis} = 0&  \label{G10}\\
& \xPickTop{\aisle}{\article}^p \in \{0,1\} & \aisle \in \setOfAisles, \article \in \setOfArticlesInAisle{\aisle} \label{G11}\\
& \xPickBottom{\aisle}{\article}^p \in \{0,1\}& \aisle \in \setOfAisles, \article \in \setOfArticlesInAisle{\aisle} \label{G12}
\end{align}
Constraints~\eqref{G1} and \eqref{G2} forbid overlaps of \horizontal branch-and-pick tours with other \horizontal branch-and-pick tours and also with the tour with cart. Constraints~\eqref{G3}--\eqref{G8} define feasible conditions for starting or continuing \horizontal branch-and-pick tours. Constraints~\eqref{G9}--\eqref{G12} define the domains of the variables. 

To ensure that the carrying capacity of the picker is not exceeded, we introduce variables $\capBottom{\aisle}$ ($\capTop{\aisle}$) that keep track of the total load collected by the picker on a  \horizontal branch-and-pick tour at the moment when passing at the bottom (at the top) of aisle $\aisle$. In Figure~\ref{fig:warehouse3}, $\capTop{4} =2$ and $\capBottom{4}=1$. Let $\itemsTop{\aisle}{\article}$ ($\itemsBottom{\aisle}{\article}$) indicate the number of required items of all SKUs stored between the top (bottom) aisle and picking position $i$, i.e.,  $\itemsTop{\aisle}{\article} = \sum_{\article' \in \setOfArticlesInAisle{\aisle}: \article' \leq\article} \sum_{\myArticle \in \setOfArticles }\demandTwo{\aisle}{\article'}{\myArticle}$ and $\itemsBottom{\aisle}{\article} = \sum_{\article' \in \setOfArticlesInAisle{\aisle}: \article' \geq\article} \sum_{\myArticle \in \setOfArticles }\demandTwo{\aisle}{\article'}{\myArticle}$. Then, we add the following constraints:

\begin{align} 
\tiny
&\capTop{\aisle+1} + \sum_{\article \in \setOfArticlesInAisle{\aisle+1}} \itemsTop{\aisle+1}{\article}\,\xPickTop{\aisle+1}{\article}^p  - C (1-\xPickerTopLeft{\aisle}) \leq \capTop{\aisle} & \aisle \in \setOfAisles\setminus \{\lastAisleVis\} \label{H1} \\
&\capBottom{\aisle+1} + \sum_{\article \in \setOfArticlesInAisle{\aisle+1}} \itemsBottom{\aisle+1}{\article} \xPickBottom{\aisle+1}{\article}^p  - C (1-\xPickerBottomLeft{\aisle}) \leq \capBottom{\aisle} & \aisle \in \setOfAisles\setminus \{\lastAisleVis\} \label{H2} \\
&\capTop{\aisle} + \sum_{\article \in \setOfArticlesInAisle{\aisle}} \itemsTop{\aisle}{\article} \xPickTop{\aisle}{\article}^p  - C (1-\xPickerTopRight{\aisle}) \leq \capTop{\aisle+1} & \aisle \in \setOfAisles\setminus \{\lastAisleVis\} \label{H3} \\
&\capBottom{\aisle} + \sum_{\article \in \setOfArticlesInAisle{\aisle}} \itemsBottom{\aisle}{\article} \xPickBottom{\aisle}{\article}^p  - C (1-\xPickerBottomRight{\aisle}) \leq \capBottom{\aisle+1} & \aisle \in \setOfAisles\setminus \{\lastAisleVis\} \label{H4} \\
& \capTop{\aisle} + \sum_{\article \in \setOfArticlesInAisle{\aisle}} \itemsTop{\aisle}{\article} \xPickTop{\aisle}{\article}^p \leq C & \aisle \in \setOfAisles \label{H5} \\
 &  \capBottom{\aisle} + \sum_{\article \in \setOfArticlesInAisle{\aisle}} \itemsBottom{\aisle}{\article} \xPickBottom{\aisle}{\article}^p \leq C & \aisle \in \setOfAisles \label{H6} \\
 & \capTop{\aisle}, \capBottom{\aisle} \geq 0 & \aisle \in \setOfAisles \label{H7} 
\end{align}
Constraints~\eqref{H1}--\eqref{H4} propagate the load to the next aisle, and constraints~\eqref{H5} and \eqref{H6} restrict the load to the maximum picker capacity.

Finally, note that the described modeling approach can also be used to implement inaccessibility constraints for certain aisles when traveling with cart. 
 
\section{The Single-Block SPRP with Multiple End Depots}
\label{sec:openDepot}
Finally, we consider the single-block SPRP with multiple end depots, in which the picker does not have to return to the start depot but may select an arbitrary end depot from a set of possible candidates. The start depot is always included in the set of candidates, and additional depots can be located in both cross aisles at the entries to all picking aisles. Initially, the relevant part of the warehouse, i.e., the set of aisles $\setOfAisles$, can only be restricted to the area between the leftmost depot or required picking position and the rightmost depot or required picking position.

\begin{figure}[ht]
	\centering
  \includegraphics[width=0.8\textwidth]{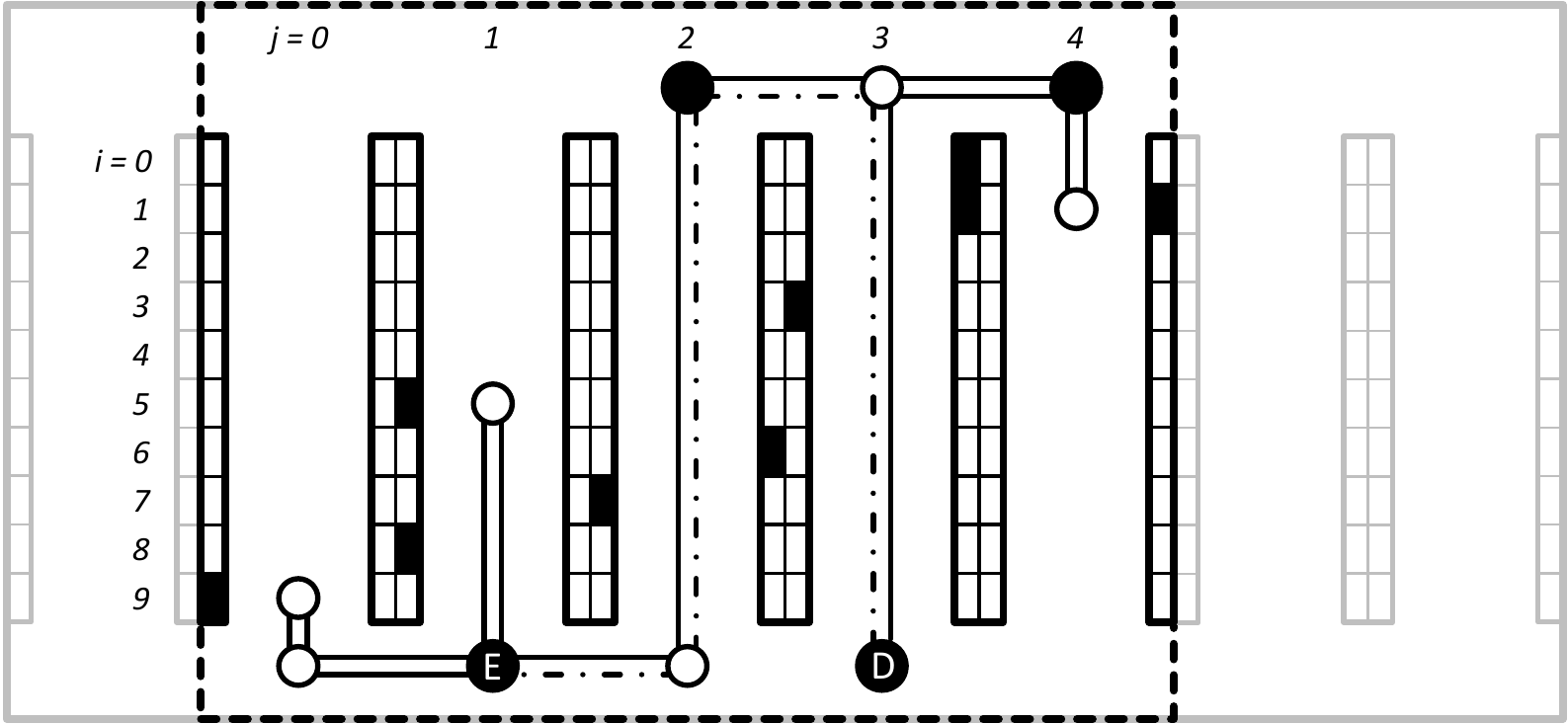}
	\caption{Optimal solution of an example instance of the single-block SPRP with multiple end depots.}
	\label{fig:warehouse4}
\end{figure}

Figure~\ref{fig:warehouse4} shows an example with the given start depot 'D' and four potential end depots given by the filled circles. Depot 'E' is finally selected as end depot. As illustrated in the figure, the idea of our modeling approach is to simultaneously determine a picking tour according to formulation~\eqref{F0}--\eqref{B5} and a simple path (given by the dot-dashed edges) on which we do not return from the selected end depot to the given start depot. The edges of this path are removed from the edges of the closed loop which starts and ends at the start depot and includes the selected end depot. To preserve the connectivity of the picking tour, the path can only contain edges that are traversed twice in the closed loop, i.e., those for which $\xTwoTopBottom{\aisle},\xTwoTop{\aisle}, \xTwoBottom{\aisle}$, or $\xTwoUp{\aisle}$ are equal to $1$. 

To define the path, we introduce the binary variables $\yTop{\aisle},\yBottom{\aisle},\yUp{\aisle}$, and $\yTopBottom{\aisle}$, which are defined analogous to the $x_{\aisle}$ variables in Section~\ref{sec:model}. The binary variables $\openDepotTop{\aisle}$ and $\openDepotBottom{\aisle}$ indicate whether the depot at the top (bottom) of aisle $\aisle$ is selected as end depot. To keep notation simple, we set $\openDepotTop{\aisle}$ and $\openDepotBottom{\aisle}$ to zero if no potential end depot is available at the respective position. The binary variables $\degreeEvenTopReturn{\aisle}$ ($\degreeEvenBottomReturn{\aisle}$) equal 0 if the degree of the path at the top (bottom) of aisle $\aisle$ is zero or uneven and equal 1 if the degree is even.
In Figure~\ref{fig:warehouse4}, the path between the start depot 'D' and the selected end depot 'E' ($\openDepotBottom{1} = 1$) is given by $\yUp{3}, \yTop{2}, \yUp{2}, \yBottom{1} = 1$. The degree of the path at the start and end depot is uneven ($\degreeEvenBottomReturn{1},\degreeEvenBottomReturn{3} = 0$), and the degrees at the top and bottom of aisle 2 and the top of aisle 3 are even ($\degreeEvenTopReturn{2}, \degreeEvenBottomReturn{2}, \degreeEvenTopReturn{3} = 1$). 

We modify the objective function \eqref{F0} and now subtract the cost associated with the path: 

\begin{align}
\tiny
& \begin{aligned}
\text{min~} &\sum_{\aisle \in \setOfAisles} \costTwoBottom{\aisle} (\xTwoBottom{\aisle}-\yBottom{\aisle}) + \costTwoTop{\aisle} (\xTwoTop{\aisle}-\yTop{\aisle}) + \costTopBottom{\aisle} \xTopBottom{\aisle}  + \costTwoTopBottom{\aisle} ( \xTwoTopBottom{\aisle}-\yTopBottom{\aisle})+  \costUp{\aisle} \xUp{\aisle} + \costTwoUp{\aisle}( \xTwoUp{\aisle}-\yUp{\aisle})+\\
&\sum_{\aisle \in \setOfAisles}\sum_{\article \in \setOfArticlesInAisle{\aisle}} \left(\costPickBottom{\aisle}{\article} \xPickBottom{\aisle}{\article} +\costPickTop{\aisle}{\article} \xPickTop{\aisle}{\article}\right)
 \end{aligned} & \label{K0}
 \end{align}
We keep constraints~\eqref{F2}--\eqref{B5} and replace constraints (2) with constraints~\eqref{F1fb}--\eqref{F1f} and constraints~\eqref{F1y} to restrict the relevant part of the warehouse: 
\begin{align} 
\tiny
&\xLast{\aisle} \geq 1 & \aisle \in \setOfAisles: \setOfArticlesInAisle{\aisle} \neq \emptyset\label{F1y}
\end{align} 
To characterize the path, we add the following constraints:
\begin{align} 
\tiny
& \xTwoTop{\aisle} +  \xTwoTopBottom{\aisle} \geq \yTop{\aisle} & \aisle \in \setOfAisles \label{K1}\\
& \xTwoBottom{\aisle} +  \xTwoTopBottom{\aisle} \geq \yBottom{\aisle}& \aisle \in \setOfAisles \label{K2}\\
& \xTwoTopBottom{\aisle} \geq \yTopBottom{\aisle}& \aisle \in \setOfAisles \label{K3}\\
& \xTwoUp{\aisle} \geq \yUp{\aisle}& \aisle \in \setOfAisles \label{K4}\\
&\yTop{\aisle} + \yBottom{\aisle+1} \leq  \yUp{\aisle+1}+1 & \aisle \in \setOfAisles \setminus \{\lastAisleVis\} \label{K6}\\
&\yBottom{\aisle} + \yTop{\aisle+1} \leq  \yUp{\aisle+1}+1 & \aisle \in \setOfAisles \setminus \{\lastAisleVis\} \label{K7}\\
& \yTop{\aisle} + \yBottom{\aisle} + \yTopBottom{\aisle} \geq  \yTop{\aisle+1} + \yBottom{\aisle+1}  + \yTopBottom{\aisle+1} & \aisle \in \setOfAisles \setminus \{\lastAisleVis\} : \aisle \geq \depotAisle \label{K9}\\
& \yTop{\aisle} + \yBottom{\aisle} +\yTopBottom{\aisle} \geq  \yTop{\aisle-1} + \yBottom{\aisle-1} +\yTopBottom{\aisle-1}  & \aisle \in \setOfAisles \setminus \{0\} : \aisle < \depotAisle \label{K10}\\
& \sum_{\aisle \in \setOfAisles} \openDepotBottom{\aisle} + \openDepotTop{\aisle} \leq 1  &  \label{K11}\\[0.1cm]
&\underset{\mathclap{j > \firstAisle}}{[}\yTopBottom{\previousAisle} +
 \yTop{\previousAisle}] +  \yTopBottom{\aisle} + 
 \yTop{\aisle}  +
\yUp{\aisle} = 2 \degreeEvenTopReturn{\aisle} +\openDepotTop{\aisle} & \begin{aligned}[c] \mathit{if}(\depotTop{} = 0)\,\{\aisle \in \setOfAisles\}\,\\\mathit{else}\,\{\aisle \in \setOfAisles \setminus \{\depotAisle\} \} \end{aligned} \label{K12}\\[0.3cm]
&\underset{\mathclap{j > \firstAisle}}{[}\yTopBottom{\previousAisle} +
 \yBottom{\previousAisle}] +  \yTopBottom{\aisle} + 
 \yBottom{\aisle}  +
\yUp{\aisle} = 2 \degreeEvenBottomReturn{\aisle} +\openDepotBottom{\aisle} & \begin{aligned}[c] \mathit{if}(\depotTop{} = 1)\,\{\aisle \in \setOfAisles\}\,\\\mathit{else}\,\{\aisle \in \setOfAisles \setminus \{\depotAisle\} \} \end{aligned}  \label{K13}\\[0.3cm]
& \yTopBottom{\aisle} + \yTop{\aisle} + \yBottom{\aisle} \leq 1& \aisle \in \setOfAisles \label{K5}\\
&\yBottom{\aisle}, \yTop{\aisle}, \yTopBottom{\aisle}, \yUp{\aisle}, \openDepotBottom{\aisle}, \openDepotTop{\aisle}, \degreeEvenTopReturn{\aisle},\degreeEvenBottomReturn{\aisle}  \in \{0,1\} & \aisle \in \setOfAisles \label{K14}
\end{align}
Constraints~\eqref{K1}--\eqref{K4} restrict the path to the edges of the closed loop that are traversed twice. Constraints~\eqref{K6}--\eqref{K10} guarantee that the path is connected. Constraint~\eqref{K11} ensures that at most one end depot is selected. Constraints \eqref{K12}--\eqref{K14} guarantee that the path is simple and connects start and end depot.

\section{Computational experiments}
\label{sec:results}
This section presents the numerical studies to assess the performance of our models and the benefits of (i)~allowing the decoupling of picker and cart, and (ii)~having multiple end depots. We discuss results for the standard SPRP (Section~\ref{sec:single}), the settings with scattered storage (Section~\ref{sec:mixed}), decoupling (Section~\ref{sec:decouple}), and multiple end depots (Section~\ref{sec:multdepot}). 

We used Gurobi\,6.5.0 set to use only a single thread and otherwise using the standard setting to solve our formulation. All studies were conducted on a Desktop PC with an AMD FX-6300 processor at 3.5 GHz with 8 GB of RAM and running Windows 10 Pro.

\subsection{Results for the Standard SPRP}
\label{sec:single}
We compare our formulation (denoted as GS) to the two best performing formulations from the literature: SHSW (\citet*{Scholz:2016}) and PPC (\citet*{Pansart:2018}). The comparison between GS and SHSW is performed on the 900-instance benchmark of \citet{Scholz:2016}, which are generated in groups of 30 instances for different numbers of aisles $\numberOfAisles \in \{5,10,15,20,25,30\}$ and required picking positions $\numberOfArticles \in \{30,45,60,75,90\}$. The number of available picking positions per aisle $\numberOfCells$ is set to 45, and the required picking positions are uniformly distributed over the warehouse. The depot is located at the bottom left of the warehouse. The comparison between GS and PPC is carried out on the benchmark of \citet{Pansart:2018}, which is generated in an analogous fashion to that of \citet{Scholz:2016} but only contains 10 instances per group. Because \citet{Pansart:2018} do not use a subset of the \citet{Scholz:2016} benchmark but generate new instances using a different random number generator, results between SHSW and PPC are not directly comparable. The two benchmarks are available at \url{http://www.mansci.ovgu.de/Forschung/Materialien/2016+_+I_-p-534.html} and \url{https://pagesperso.g-scop.grenoble-inp.fr/~pansartl/en/en_picking.html}, respectively. 

Table~\ref{tab:resultsForm} shows the results of the comparison: Column  $(\numberOfAisles,\numberOfArticles)$ denotes the instance group defined as combination of the number of aisles $\numberOfAisles$ and the number of required picking positions $\numberOfArticles$. Column $\# \solvedInst$ reports the number of instances solved to optimality with each formulation within the runtime limit of 30 minutes.  In column \runtime{avg}, we report the average runtime in seconds calculated over all instances in the respective group that were solved by the respective method within the time limit, and in column \runtime{max} the maximum runtime observed for any of the instances in the considered group.
We judge the comparison of runtimes to be relatively fair: the speed of our processor and the one used by \citet{Pansart:2018} (an Intel Xeon E5-2440 v2 at 1.9\,GHz) are roughly comparable based on their Passmark single-thread scores; \citet{Scholz:2016} only provide the information that a 3.4\,GHz Pentium processor was used, but in general, models that fit this description have similar or even superior Passmark single-thread scores than the other two machines.

\begin{table}[htbp]
\centering
\begin{footnotesize}
\sffamily
\setlength{\tabcolsep}{5.0pt}
   \ra{1.1} \begin{tabular}{@{}l@{\hspace{0.5cm}}rr@{\hspace{0.5cm}}rrr@{\hspace{.5cm}}rrr@{\hspace{0.5cm}}rrr@{}} \toprule
      & \multicolumn{2}{c}{SHSW} & \multicolumn{3}{c}{GS}   & \multicolumn{3}{c}{PPC} & \multicolumn{3}{c}{GS}\\ \cmidrule(rr{.55cm}l{.05cm}){2-3} \cmidrule(r{.05cm}l{.05cm}){4-6} \cmidrule(r{.05cm}l{.05cm}){7-9} \cmidrule(r{.05cm}l{.05cm}){10-12}
    $(\numberOfAisles,\numberOfArticles)$ &\multicolumn{1}{c}{\#\solvedInst.} & \multicolumn{1}{c}{\runtime{avg}} & \multicolumn{1}{c}{\#\solvedInst.} & \multicolumn{1}{c}{\runtime{avg}} & \runtime{max} & \multicolumn{1}{c}{\#\solvedInst.} & \multicolumn{1}{c}{\runtime{avg}} & \multicolumn{1}{c}{\runtime{max}} & \multicolumn{1}{c}{\#\solvedInst.} & \multicolumn{1}{c}{\runtime{avg}} & \multicolumn{1}{c}{\runtime{max}} \\ 
\midrule
(5,30) & 30/30 & 0.09  & 30/30 & 0.01  & 0.02  & 10/10 & 0.03  & 0.08  & 10/10 & 0.00  & 0.02 \\
(5,45) & 30/30 & 0.09  & 30/30 & 0.01  & 0.02  & 10/10 & 0.06  & 0.09  & 10/10 & 0.01  & 0.02 \\
(5,60) & 30/30 & 0.09  & 30/30 & 0.01  & 0.02  & 10/10 & 0.08  & 0.10  & 10/10 & 0.02  & 0.03 \\
(5,75) & 30/30 & 0.09  & 30/30 & 0.01  & 0.02  & 10/10 & 0.08  & 0.11  & 10/10 & 0.02  & 0.03 \\
(5,90) & 30/30 & 0.10  & 30/30 & 0.01  & 0.02  & 10/10 & 0.08  & 0.13  & 10/10 & 0.03  & 0.03 \\ \addlinespace
 (10,30) & 30/30 & 1.60  & 30/30 & 0.02  & 0.05  & 10/10 & 0.05  & 0.13  & 10/10 & 0.03  & 0.06 \\
(10,45) & 30/30 & 1.03  & 30/30 & 0.02  & 0.03  & 10/10 & 0.08  & 0.19  & 10/10 & 0.03  & 0.05 \\
(10,60) & 30/30 & 1.42  & 30/30 & 0.02  & 0.05  & 10/10 & 0.13  & 0.20  & 10/10 & 0.02  & 0.05 \\
(10,75) & 30/30 & 1.36  & 30/30 & 0.02  & 0.05  & 10/10 & 0.09  & 0.17  & 10/10 & 0.02  & 0.05 \\
(10,90) & 30/30 & 0.62  & 30/30 & 0.02  & 0.03  & 10/10 & 0.10  & 0.20  & 10/10 & 0.02  & 0.05 \\ \addlinespace
(15,30) & 30/30 & 2.29  & 30/30 & 0.02  & 0.10  & 10/10 & 0.06  & 0.13  & 10/10 & 0.04  & 0.08 \\
(15,45) & 30/30 & 5.28  & 30/30 & 0.03  & 0.06  & 10/10 & 0.11  & 0.20  & 10/10 & 0.04  & 0.05 \\
(15,60) & 30/30 & 10.64 & 30/30 & 0.03  & 0.08  & 10/10 & 0.14  & 0.32  & 10/10 & 0.05  & 0.08 \\
(15,75) & 30/30 & 15.10 & 30/30 & 0.03  & 0.08  & 10/10 & 0.15  & 0.41  & 10/10 & 0.04  & 0.08 \\
(15,90) & 30/30 & 19.41 & 30/30 & 0.04  & 0.08  & 10/10 & 0.23  & 0.45  & 10/10 & 0.05  & 0.08 \\\addlinespace
(20,30) & 30/30 & 10.57 & 30/30 & 0.04  & 0.16  & 10/10 & 0.09  & 0.31  & 10/10 & 0.06  & 0.08 \\
(20,45) & 30/30 & 27.32 & 30/30 & 0.03  & 0.13  & 10/10 & 0.10  & 0.21  & 10/10 & 0.07  & 0.11 \\
(20,60) & 30/30 & 114.33 & 30/30 & 0.04  & 0.08  & 10/10 & 0.26  & 0.49  & 10/10 & 0.06  & 0.11 \\
(20,75) & 30/30 & 216.63 & 30/30 & 0.04  & 0.08  & 10/10 & 0.84  & 5.21  & 10/10 & 0.07  & 0.13 \\
(20,90) & 30/30 & 485.71 & 30/30 & 0.05  & 0.11  & 10/10 & 0.39  & 1.64  & 10/10 & 0.09  & 0.13 \\\addlinespace
(25,30) & 30/30 & 54.46 & 30/30 & 0.04  & 0.14  & 10/10 & 0.11  & 0.24  & 10/10 & 0.05  & 0.06 \\
(25,45) & 30/30 & 85.46 & 30/30 & 0.05  & 0.13  & 10/10 & 0.24  & 0.50  & 10/10 & 0.06  & 0.10 \\
(25,60) & 30/30 & 258.92 & 30/30 & 0.06  & 0.13  & 10/10 & 0.71  & 2.34  & 10/10 & 0.08  & 0.17 \\
(25,75) & 29/30 & 527.39 & 30/30 & 0.07  & 0.19  & 10/10 & 0.76  & 2.34  & 10/10 & 0.09  & 0.15 \\
(25,90) & 24/30 & 646.59 & 30/30 & 0.07  & 0.18  & 10/10 & 1.01  & 4.41  & 10/10 & 0.09  & 0.18 \\\addlinespace
(30,30) & 30/30 & 204.18 & 30/30 & 0.05  & 0.11  & 10/10 & 0.08  & 0.21  & 10/10 & 0.06  & 0.13 \\
(30,45) & 30/30 & 406.19 & 30/30 & 0.06  & 0.17  & 10/10 & 0.19  & 0.49  & 10/10 & 0.08  & 0.11 \\
(30,60) & 30/30 & 508.80 & 30/30 & 0.07  & 0.17  & 10/10 & 0.39  & 0.59  & 10/10 & 0.10  & 0.13 \\
(30,75) & 24/30 & 638.89 & 30/30 & 0.07  & 0.16  & 10/10 & 0.99  & 6.69  & 10/10 & 0.11  & 0.19 \\
(30,90) & 21/30 & 786.29 & 30/30 & 0.08  & 0.22  & 10/10 & 1.63  & 6.69  & 10/10 & 0.15  & 0.30 \\
\midrule
\textbf{Avg.} &  & \textbf{167.70} & & \textbf{0.04} & & & \textbf{0.31} &  & & \textbf{0.05} &  \\
 \bottomrule
    \end{tabular}%
\rmfamily
  \caption{\textrm{Comparison of the formulations SHSW, PPC, and GS on standard SPRP instances from the literature}.\label{tab:resultsForm}}
    \end{footnotesize}
   \end{table}

Comparing GS to SHSW, we note that GS is able to solve all instances, while SHSW fails to solve 22 of the instances with a number of aisles $\numberOfAisles \geq 25$ and a number of required picking positions $\numberOfArticles \geq 75$. The runtimes of SHSW grow strongly with a larger number of aisles, and, for a number of aisles above 10 also with an increasing number of required picking positions. Contrary to this, the runtime of GS only grows moderately with a larger number of aisles, and the number of required picking positions seems to have no clear influence on the runtime. The average runtime of GS is approximately 4500 times lower than that of SHSW. The difference between GS and PPC is smaller: Both formulations are able to solve all instance groups within less than 10 seconds. Still, GS is approximately six times faster on average, and the worst runtime on the instance groups is up to 40 times lower than that of PPC.

To assess the scaling behavior of GS, we generate a set of large instances, more precisely, 30 instances for all combinations of $\numberOfAisles, \numberOfArticles, \numberOfCells \in \{100,250,500,750,1000\}$. Again, the required picking positions are uniformly distributed over the warehouse, and the depot is located at the bottom left corner of the warehouse. Table~\ref{tab:resultsLarge} presents aggregate results over the instances in a group. In addition to the values reported in Table~\ref{tab:resultsForm}, we report in rows \runtimeSup{avg}{\numberOfArticles} the average runtimes over different values of $\numberOfArticles$ within one group (defined by a fixed value of $\numberOfAisles$ and $\numberOfCells$). Column \runtimeSup{avg}{\numberOfCells} reports the average runtime over all different values of $\numberOfCells$ for a fixed combination of $\numberOfAisles$ and $\numberOfArticles$.

\begin{table}[htbp]
  \centering
\begin{footnotesize}
\sffamily
\setlength{\tabcolsep}{5.0pt}
   \ra{1.1} \begin{tabular}{@{}l@{\hspace{0.5cm}}rr@{\hspace{0.5cm}}rr@{\hspace{0.5cm}}rr@{\hspace{0.5cm}}rr@{\hspace{0.5cm}}rr@{\hspace{0.5cm}}r@{}}\toprule
     & \multicolumn{2}{c}{$\numberOfCells = 100$} & \multicolumn{2}{c}{$\numberOfCells = 250$} & \multicolumn{2}{c}{$\numberOfCells = 500$} & \multicolumn{2}{c}{$\numberOfCells = 750$} & \multicolumn{2}{c}{$\numberOfCells = 1000$} \\ \cmidrule(l{.2cm}r{.2cm}){2-3} \cmidrule(l{.2cm}r{.2cm}){4-5} \cmidrule(l{.2cm}r{.2cm}){6-7} \cmidrule(l{.2cm}r{.2cm}){8-9} \cmidrule(l{.2cm}r{.2cm}){10-11}
 $(\numberOfAisles,\numberOfArticles)$     & \multicolumn{1}{r}{\runtime{avg}} & \multicolumn{1}{r@{\hspace{0.5cm}}}{\runtime{max}} & \multicolumn{1}{r}{\runtime{avg}} & \multicolumn{1}{r@{\hspace{0.5cm}}}{\runtime{max}} & \multicolumn{1}{r}{\runtime{avg}} & \multicolumn{1}{r@{\hspace{0.5cm}}}{\runtime{max}} & \multicolumn{1}{r}{\runtime{avg}} & \multicolumn{1}{r@{\hspace{0.5cm}}}{\runtime{max}} & \multicolumn{1}{r}{\runtime{avg}} & \multicolumn{1}{r}{\runtime{max}} & \multicolumn{1}{r}{\boldmath\runtimeSup{avg}{\numberOfCells}} \\\midrule
 
(100,100) & 1.03  & 2.38  & 1.33  & 5.50  & 1.87  & 13.47 & 1.23  & 6.49  & 1.30  & 4.63  & \textbf{1.35} \\
(100,250) & 1.67  & 8.87  & 2.08  & 7.11  & 1.79  & 5.06  & 1.54  & 4.82  & 2.22  & 6.44  & \textbf{1.86} \\
(100,500) & 2.50  & 10.20 & 1.35  & 4.49  & 1.30  & 4.60  & 1.13  & 2.13  & 1.14  & 1.87  & \textbf{1.49} \\
(100,750) & 1.38  & 5.09  & 1.32  & 3.81  & 1.65  & 6.05  & 1.42  & 6.29  & 1.16  & 1.57  & \textbf{1.39} \\
(100,1000) & 1.46  & 6.12  & 1.65  & 4.80  & 1.27  & 2.79  & 1.27  & 2.87  & 1.45  & 2.86  & \textbf{1.42} \\
{\boldmath\runtimeSup{avg}{\numberOfArticles}}     & \textbf{1.61} &       & \textbf{1.54} &       & \textbf{1.58} &       & \textbf{1.32} &       & \textbf{1.45} &       &  \\
\midrule
(250,100) & 2.42  & 5.21  & 3.58  & 17.52 & 3.43  & 12.50 & 5.60  & 23.89 & 4.79  & 32.91 & \textbf{3.96} \\
(250,250) & 5.55  & 22.54 & 7.52  & 24.61 & 10.98 & 29.25 & 9.93  & 28.51 & 9.14  & 26.16 & \textbf{8.62} \\
(250,500) & 8.26  & 29.42 & 6.07  & 32.26 & 10.10 & 26.03 & 7.59  & 26.32 & 8.97  & 29.03 & \textbf{8.20} \\
(250,750) & 11.06 & 44.55 & 12.77 & 33.40 & 14.10 & 36.57 & 12.35 & 30.44 & 14.21 & 35.33 & \textbf{12.90} \\
(250,1000) & 14.99 & 34.76 & 15.03 & 35.66 & 9.96  & 32.23 & 11.32 & 31.41 & 9.22  & 25.84 & \textbf{12.10} \\
{\boldmath\runtimeSup{avg}{\numberOfArticles}}   & \textbf{8.45} &       & \textbf{8.99} &       & \textbf{9.71} &       & \textbf{9.36} &       & \textbf{9.27} &       &  \\
\midrule     
(500,100) & 4.32  & 15.45 & 6.78  & 18.43 & 8.82  & 31.58 & 11.46 & 47.96 & 15.22 & 68.86 & \textbf{9.32} \\
(500,250) & 14.44 & 63.67 & 12.81 & 37.21 & 22.07 & 68.54 & 28.85 & 54.87 & 26.87 & 87.59 & \textbf{21.01} \\
(500,500) & 11.43 & 56.09 & 18.99 & 66.95 & 23.75 & 59.27 & 29.56 & 64.88 & 34.35 & 75.92 & \textbf{23.62} \\
(500,750) & 17.86 & 55.94 & 21.56 & 66.74 & 26.19 & 71.22 & 25.35 & 68.96 & 30.05 & 79.15 & \textbf{24.20} \\
(500,1000) & 16.03 & 64.74 & 18.64 & 62.14 & 23.48 & 71.43 & 30.40 & 83.83 & 24.12 & 63.57 & \textbf{22.53} \\
{\boldmath\runtimeSup{avg}{\numberOfArticles}}   & \textbf{12.81} &       & \textbf{15.75} &       & \textbf{20.86} &       & \textbf{25.12} &       & \textbf{26.13} &       &  \\
\midrule     
(750,100) & 8.68  & 41.81 & 8.70  & 55.56 & 12.67 & 77.53 & 16.09 & 114.41 & 16.18 & 117.38 & \textbf{12.46} \\
(750,250) & 19.75 & 56.34 & 29.74 & 95.73 & 44.66 & 107.42 & 55.18 & 125.61 & 50.61 & 120.86 & \textbf{39.99} \\
(750,500) & 24.11 & 73.18 & 56.25 & 134.75 & 64.34 & 161.30 & 72.61 & 211.40 & 68.02 & 166.36 & \textbf{57.06} \\
(750,750) & 31.09 & 112.22 & 44.27 & 156.87 & 60.81 & 140.97 & 68.15 & 130.10 & 68.59 & 154.34 & \textbf{54.58} \\
(750,1000) & 23.09 & 106.81 & 34.24 & 104.50 & 50.49 & 130.80 & 52.78 & 127.16 & 72.75 & 173.93 & \textbf{46.67} \\
{\boldmath\runtimeSup{avg}{\numberOfArticles}}      & \textbf{21.34} &       & \textbf{34.64} &       & \textbf{46.59} &       & \textbf{52.96} &       & \textbf{55.23} &       &  \\
\midrule 
(1000,100) & 10.40 & 74.06 & 26.68 & 300.07 & 22.36 & 98.25 & 23.42 & 141.61 & 22.77 & 154.93 & \textbf{21.12} \\
(1000,250) & 24.25 & 84.93 & 47.38 & 174.19 & 74.49 & 257.29 & 82.95 & 371.27 & 89.06 & 617.04 & \textbf{63.63} \\
(1000,500) & 25.07 & 72.91 & 70.79 & 227.91 & 83.98 & 181.89 & 110.82 & 287.68 & 90.04 & 291.63 & \textbf{76.14} \\
(1000,750) & 33.60 & 127.68 & 117.88 & 310.40 & 161.87 & 338.99 & 122.89 & 239.56 & 102.21 & 242.78 & \textbf{107.69} \\
(1000,1000) & 35.62 & 134.22 & 96.56 & 261.85 & 127.16 & 307.27 & 106.87 & 226.21 & 122.96 & 251.78 & \textbf{97.83} \\
{\boldmath\runtimeSup{avg}{\numberOfArticles}}      & \textbf{25.79} &       & \textbf{71.86} &       & \textbf{93.97} &       & \textbf{89.39} &       & \textbf{85.41} &       &  \\
\bottomrule
    \end{tabular}%
     \end{footnotesize}
\rmfamily
  \caption{Results of our formulation on newly generated large standard SPRP instances\label{tab:resultsLarge}.}%
\end{table}%

The results show that the runtimes consistently increase for a larger number of aisles $\numberOfAisles$, while the relationship between the number of required picking positions $\numberOfArticles$ or the number of available picking positions  $\numberOfCells$ and the runtime is not entirely consistent: we can only note the rough tendency that runtimes are higher for larger values of the two parameters. Overall, the scaling behavior of our formulation is quite convincing, even the largest instances with $\numberOfAisles, \numberOfArticles, \numberOfCells = 1000$ can be solved with an average runtime of about two minutes, and the most challenging instance in the benchmark was solved in approximately 10 minutes.

\subsection{Results for the Single-Block SPRP with Scattered Storage}
\label{sec:mixed}
Because the benchmark instances used by \citet{Weidinger:2018a} were not archived by the author, we generated new instances in a fashion similar to the procedure described in the original paper. To allow for a fair comparison, we reimplemented the mathematical model of \citet{Weidinger:2018a} using Gurobi (in the following denoted as formulation W). 

The instances consider warehouses of different sizes by varying the number of aisles $\numberOfAisles \in \{5,25,100\}$ and the number of available picking positions $\numberOfCells \in \{30,60,180\}$. We assume that one SKU is stored in each picking position. To investigate the influence of the degree of duplication of the SKUs in the warehouse, we vary the number of different SKUs $\xi$ stored in the warehouse depending on (i)~the storage capacity of the warehouse (given by $\numberOfAisles \cdot \numberOfCells$), (ii)~a factor $\alpha \in \{1,5,10,50\}$ that determines the frequency with which SKUs are assigned to multiple storage positions, and (iii)~the number of different SKUs in the pick list $\numberOfArticles$ as follows: 
$$ \xi =\max(\numberOfArticles,\lceil \numberOfAisles \cdot \numberOfCells / \alpha \rceil).$$ For example, if we set  $\alpha = 1$, we have a standard warehouse in which each picking position is occupied by a different SKU. The maximum expression guarantees that for higher degrees of duplication at least as many different SKUs as required in the pick list are available in the warehouse. 

The SKUs in the warehouse are divided into three classes A, B, and C based on their turnover rate. We assign $20\%$ of the SKUs to class A, $30\%$ to class B, and $50\%$ to class C. To ensure that each SKU is available at least at one picking position, each SKU is first assigned to one randomly selected picking position. Afterward, all remaining picking positions are assigned a randomly drawn SKU from class A with $80\%$ probability, from class B with $15\%$ probability, and from class C with 5\% probability. The number of items of the selected SKU that is available at each picking position is randomly selected from $\mathbb{N} \cap [1,3]$. 
Next, we generate pick lists with $\numberOfArticles \in \{3,7,15,30\}$ SKUs. Each SKU is selected from classes A, B, and C according to the probabilities $80\%$, $15\%$, and $5\%$. The demand $\demand{\myArticle}$ for each SKU $\myArticle$ is randomly drawn from $\mathbb{N} \cap [1,min(6,\bar{\capacity{\myArticle}{}})]$ with $\bar{\capacity{\myArticle}{}}$ the total supply of $\myArticle$. In the described way, we generate $3 \cdot 3 \cdot 4 \cdot 4 = 144$ instances. 

Both formulations were given a time limit of one hour. Table~\ref{tab:mixed} reports the results for different warehouse sizes $(m,n)$, number of SKUs in the pick list $a$, and degrees of duplication $\alpha$. For W, we report the runtime in column $t_W$ (TL indicates that the time limit was reached, OOM that an out of memory error occurs) and the difference between the upper bound found and the optimal solution (if no valid upper bound is found, this is indicated with a ``-''). GS finds the optimal solution for all instances, and we only report the runtime in column $t_{GS}$. If all instances of an instance group were solved to optimality by the respective formulation, we report averages of the runtimes for the group.

\begin{table}[htbp]
  \centering
    \begin{footnotesize}
\sffamily
\setlength{\tabcolsep}{3.0pt}
   \ra{1.1} \begin{tabular}{@{}l@{\hspace{0.6cm}}rr@{\hspace{0.5cm}}r@{\hspace{0.9cm}}rr@{\hspace{0.5cm}}r@{\hspace{0.9cm}}rr@{\hspace{0.5cm}}r@{\hspace{0.9cm}}rr@{\hspace{0.5cm}}r@{}}
    \toprule
          & \multicolumn{3}{c}{$a=3$} & \multicolumn{3}{c}{$a=7$} & \multicolumn{3}{c}{$a=15$} & \multicolumn{3}{c}{$a=30$} \\ \cmidrule(l{.0cm}r{.5cm}){2-4} \cmidrule(l{.0cm}r{.5cm}){5-7} \cmidrule(l{.0cm}r{.5cm}){8-10} \cmidrule(l{.0cm}r{.0cm}){11-13}
    $(m,n)$ & $\Delta_{ub}$ & $t_{W}$ & $t_{GS}$   & $\Delta_{ub}$ & $t_{W}$ & $t_{GS}$   & $\Delta_{ub}$ & $t_{W}$ & $t_{GS}$   & $\Delta_{ub}$ & $t_{W}$ & $t_{GS}$ \\
    \midrule
{\boldmath$\alpha=1$} &       &       &       &       &       &       &       &       &       &       &       &  \\
(5,30) & 0.0   & 0.00  & 0.01  & 0.0   & 0.02  & 0.00  & 0.0   & 0.88  & 0.01  & 0.0   & 98.92 & 0.02 \\
(5,60) & 0.0   & 0.00  & 0.01  & 0.0   & 0.01  & 0.00  & 0.0   & 0.93  & 0.01  & 0.0   & 3.73  & 0.01 \\
(5,180) & 0.0   & 0.03  & 0.00  & 0.0   & 0.04  & 0.01  & 0.0   & 0.14  & 0.01  & 0.0   & 14.41 & 0.01 \\
(25,30) & 0.0   & 0.01  & 0.08  & 0.0   & 0.03  & 0.14  & 0.0   & 3.04  & 0.07  & 0.0   & 570.42 & 0.14 \\
(25,60) & 0.0   & 0.01  & 0.03  & 0.0   & 0.03  & 0.02  & 0.0   & 9.98  & 0.15  & 0.0   & TL    & 0.05 \\
(25,180) & 0.0   & 0.01  & 0.15  & 0.0   & 0.04  & 0.06  & 0.0   & 0.23  & 0.03  & 0.0   & 114.35 & 0.10 \\
(100,30) & 0.0   & 0.01  & 0.21  & 0.0   & 0.03  & 0.93  & 0.0   & 56.99 & 0.15  & 0.4   & TL    & 0.15 \\
(100,60) & 0.0   & 0.00  & 0.53  & 0.0   & 0.03  & 0.28  & 0.0   & 15.22 & 0.12  & 0.2   & TL    & 0.44 \\
(100,180) & 0.0   & 0.01  & 0.67  & 0.0   & 0.04  & 0.30  & 0.0   & 0.23  & 0.13  & 0.0   & 1761.86 & 0.17 \\
\textbf{Avg.} &       & \textbf{0.01} & \textbf{0.19} &       & \textbf{0.03} & \textbf{0.19} &       & \textbf{9.74} & \textbf{0.07} &       &       & \textbf{0.12} \\
\midrule
{\boldmath$\alpha=5$} &       &       &       &       &       &       &       &       &       &       &       &  \\
(5,30) & 0.0   & 1.09  & 0.03  & 0.0   & TL    & 0.05  & 1.2   & TL    & 0.10  & 16.0  & TL    & 0.06 \\
(5,60) & 0.0   & 2.63  & 0.05  & 0.0   & TL    & 0.02  & 6.8   & TL    & 0.07  & 86.9  & TL    & 0.10 \\
(5,180) & 0.0   & 1.54  & 0.01  & 12.3  & TL    & 0.52  & 106.1 & TL    & 0.18  & -     & TL    & 0.75 \\
(25,30) & 0.0   & TL    & 0.39  & 3.9   & TL    & 0.20  & 86.1  & TL    & 0.16  & 320.3 & TL    & 0.16 \\
(25,60) & 0.0   & 1.59  & 0.10  & 3.4   & TL    & 0.31  & 34.7  & TL    & 0.98  & -     & TL    & 2.24 \\
(25,180) & 0.0   & 2.68  & 0.25  & 0.0   & 128.34 & 0.07  & 26.5  & TL    & 0.39  & -     & TL    & 1.59 \\
(100,30) & 0.0   & TL    & 0.27  & 15.3  & TL    & 0.19  & 72.8  & TL    & 1.99  & -     & TL    & 3.18 \\
(100,60) & 0.0   & 28.33 & 0.30  & 1.6   & TL    & 1.24  & 13.7  & TL    & 0.55  & -     & TL    & 0.48 \\
(100,180) & 0.0   & 15.84 & 3.37  & 12.1  & TL    & 1.72  & 148.5 & TL    & 1.83  & 105.8 & TL    & 1.13 \\
\textbf{Avg.} &       &       & \textbf{0.53} &       &       & \textbf{0.48} &       &       & \textbf{0.69} &       &       & \textbf{1.08} \\\midrule
{\boldmath$\alpha=10$} &       &       &       &       &       &       &       &       &       &       &       &  \\
(5,30) & 0.0   & TL    & 0.02  & 0.0   & TL    & 0.04  & 3.0   & TL    & 0.05  & 14.9  & TL    & 0.02 \\
(5,60) & 0.0   & 17.35 & 0.02  & 0.0   & TL    & 0.20  & 83.0  & TL    & 0.07  & 37.8  & TL    & 0.19 \\
(5,180) & 0.0   & TL    & 0.07  & 68.9  & TL    & 0.32  & 77.0  & TL    & 0.36  & -     & TL    & 0.64 \\
(25,30) & 0.0   & 92.49 & 0.19  & 10.7  & TL    & 0.33  & 28.6  & TL    & 0.72  & -     & TL    & 0.27 \\
(25,60) & 0.0   & 447.62 & 4.24  & 32.7  & TL    & 0.44  & 375.0 & TL    & 1.00  & -     & TL    & 0.87 \\
(25,180) & 0.0   & 86.79 & 0.47  & 74.7  & TL    & 0.36  & 525.8 & TL    & 30.94 & -     & TL    & 3.11 \\
(100,30) & 0.0   & TL    & 0.51  & 12.6  & TL    & 0.82  & 412.1 & TL    & 1.43  & -     & TL    & 2.79 \\
(100,60) & 0.0   & 418.16 & 0.33  & 15.9  & TL    & 11.69 & 310.9 & TL    & 3.21  & -     & TL    & 6.04 \\
(100,180) & 0.0   & 849.48 & 1.69  & 38.3  & TL    & 22.58 & -     & TL    & 93.84 & -     & OOM   & 29.07 \\
\textbf{Avg.} &       &       & \textbf{0.84} &       &       & \textbf{4.09} &       &       & \textbf{14.62} &       &       & \textbf{4.78} \\\midrule
{\boldmath$\alpha=40$} &       &       &       &       &       &       &       &       &       &       &       &  \\
(5,30) & 0.0   & TL    & 0.38  & 2.1   & TL    & 0.19  & 0.0   & TL    & 0.03  & 7.1   & TL    & 0.04 \\
(5,60) & 0.0   & TL    & 0.39  & 50.9  & TL    & 1.00  & 20.6  & TL    & 0.12  & 84.5  & TL    & 0.10 \\
(5,180) & 0.0   & TL    & 0.76  & 333.9 & TL    & 2.07  & -     & TL    & 1.26  & -     & TL    & 0.74 \\
(25,30) & 0.0   & TL    & 0.12  & 607.9 & TL    & 0.35  & -     & TL    & 1.28  & -     & TL    & 0.49 \\
(25,60) & 29.2  & TL    & 0.47  & -     & TL    & 0.68  & -     & TL    & 4.74  & -     & TL    & 5.31 \\
(25,180) & 178.0 & TL    & 0.88  & -     & OOM   & 0.79  & -     & TL    & 1.20  & -     & TL    & 8.98 \\
(100,30) & 439.5 & TL    & 0.78  & -     & OOM   & 2.41  & -     & TL    & 3.25  & -     & OOM   & 4.90 \\
(100,60) & 0.0   & 193.97 & 0.11  & -     & OOM   & 81.39 & -     & TL    & 21.61 & -     & TL    & 12.63 \\
(100,180) & 0.0   & 245.70 & 1.76  & -     & OOM   & 1.30  & -     & OOM   & 4.75  & -     & OOM   & 155.13 \\
\textbf{Avg.} &       &       & \textbf{0.63} &       &       & \textbf{10.02} &       &       & \textbf{4.25} &       &       & \textbf{20.92} \\
\bottomrule
    \end{tabular}%
    \end{footnotesize}
\rmfamily
	\caption{Comparison of formulations W and GS for the single-block SPRP with scattered storage.}
  \label{tab:mixed}%
\end{table}%

For W, both a higher number of SKUs in the pick list and a higher degree of duplication make the instances more difficult to solve. Of the 144 instances, 95 cannot be solved to proven optimality within the time limit (no valid upper bound is found in 24 cases, and an OOM error occurs in 8 cases). Contrary to this, GS is able to solve all instances, and the average runtime on the hardest instance group is approximately 21 seconds. The highest runtime observed for any instance is around 155 seconds. The relationship between the number of SKUs in the pick list or the degree of duplication and the runtime of GS is not entirely consistent: we can again note only a rough tendency that runtimes are higher for larger values of the two parameters.

\subsection{Results for the Single-Block SPRP with Decoupling}
\label{sec:decouple}
To investigate the performance of our formulation in the setting with decoupling of picker and cart, we use the standard SPRP benchmark of \citet{Scholz:2016}. Because we want to study the effect of different capacities and speeds of the picker when traveling without cart, we consider three values of the picker capacity $C \in \{2,4,6\}$, and we vary the travel time required by the picker without cart by multiplying the original travel time with cart by factors $\beta\in\{0.5,0.75\}$, e.g., $\costTwoTop{\aisle}^p = \beta\cdot \costTwoTop{\aisle}$. Thus, we study $6 \cdot 900 = 5400$ instances. Table~\ref{tab:decouple} provides aggregate values for the 30 instances in each group $(\numberOfAisles,\numberOfArticles)$: the average gap $\Delta$ between the objective value obtained with the respective setting and the base setting without decoupling of picker and cart, the average runtime \runtime{avg}, and the maximum runtime \runtime{max}. In the last row, we provide the average of the gaps and of the average runtimes for all combinations of travel speed and capacity of the picker.  

\afterpage{
\begin{landscape}
\begin{table}[p]
\centering
\ra{1.1}
 \begin{footnotesize}
\sffamily
\setlength{\tabcolsep}{2.0pt}
\begin{tabular}{@{}l@{\hspace{0.5cm}}rrr@{\hspace{0.5cm}}rrr@{\hspace{0.5cm}}rrr@{\hspace{0.5cm}}rrr@{\hspace{0.5cm}}rrr@{\hspace{0.5cm}}rrr@{}}
\toprule
 & \multicolumn{6}{c}{$C = 2$} & \multicolumn{6}{c}{$C = 4$} & \multicolumn{6}{c}{$C = 6$} \\  \cmidrule(l{.2cm}r{.2cm}){2-7} \cmidrule(l{.2cm}r{.2cm}){8-13} \cmidrule(l{.2cm}r{.0cm}){14-19}  & \multicolumn{3}{c}{$\beta = 0.75$} & \multicolumn{3}{c}{$\beta = 0.5$}& \multicolumn{3}{c}{$\beta = 0.75$} & \multicolumn{3}{c}{$\beta = 0.5$} & \multicolumn{3}{c}{$\beta = 0.75$} & \multicolumn{3}{c}{$\beta = 0.5$} \\  \cmidrule(l{.2cm}r{.2cm}){2-4} \cmidrule(l{.0cm}r{.2cm}){5-7} \cmidrule(l{.2cm}r{.2cm}){8-10} \cmidrule(l{.2cm}r{.2cm}){11-13} \cmidrule(l{.2cm}r{.2cm}){14-16} \cmidrule(l{.2cm}r{.0cm}){17-19}
$(\numberOfAisles,\numberOfArticles)$  & $\Delta$ & \runtime{avg} & \runtime{max}  & $\Delta$ & \runtime{avg} & \runtime{max}  & $\Delta$ & \runtime{avg} & \runtime{max} &  $\Delta$ & \runtime{avg} & \runtime{max} & $\Delta$ & \runtime{avg} & \runtime{max} & $\Delta$ & \runtime{avg} & \runtime{max} \\
\midrule
(5,30) & -2.14 & 0.01  & 0.03  & -6.01 & 0.01  & 0.02  & -4.39 & 0.02  & 0.03  & -17.02 & 0.03  & 0.12  & -5.77 & 0.02  & 0.03  & -20.51 & 0.13  & 0.22 \\
(5,45) & -1.88 & 0.02  & 0.03  & -4.25 & 0.01  & 0.03  & -4.00 & 0.02  & 0.03  & -11.36 & 0.02  & 0.06  & -5.03 & 0.02  & 0.03  & -17.52 & 0.07  & 0.22 \\
(5,60) & -1.40 & 0.02  & 0.05  & -3.14 & 0.02  & 0.05  & -3.17 & 0.03  & 0.06  & -7.11 & 0.02  & 0.05  & -4.66 & 0.03  & 0.06  & -13.06 & 0.03  & 0.09 \\
(5,75) & -0.96 & 0.03  & 0.05  & -2.19 & 0.03  & 0.05  & -2.87 & 0.03  & 0.05  & -6.66 & 0.03  & 0.05  & -4.30 & 0.03  & 0.05  & -9.80 & 0.03  & 0.06 \\
(5,90) & -0.79 & 0.04  & 0.05  & -1.66 & 0.03  & 0.05  & -2.03 & 0.03  & 0.06  & -4.51 & 0.03  & 0.06  & -3.36 & 0.03  & 0.06  & -7.53 & 0.03  & 0.05 \\\addlinespace
(10,30) & -7.43 & 0.06  & 0.14  & -18.24 & 0.15  & 0.62  & -12.25 & 0.11  & 0.37  & -29.06 & 0.55  & 2.37  & -12.95 & 0.11  & 0.37  & -30.23 & 1.19  & 2.68 \\
(10,45) & -3.83 & 0.05  & 0.09  & -10.47 & 0.08  & 0.28  & -7.31 & 0.07  & 0.26  & -23.85 & 0.46  & 1.95  & -8.40 & 0.07  & 0.26  & -27.80 & 1.66  & 5.08 \\
(10,60) & -2.26 & 0.05  & 0.13  & -6.81 & 0.07  & 0.22  & -5.59 & 0.07  & 0.18  & -20.25 & 0.28  & 3.77  & -6.89 & 0.07  & 0.18  & -25.97 & 0.71  & 4.23 \\
(10,75) & -1.30 & 0.04  & 0.06  & -3.73 & 0.06  & 0.16  & -3.27 & 0.06  & 0.11  & -13.92 & 0.13  & 0.36  & -4.48 & 0.06  & 0.11  & -21.74 & 0.36  & 2.06 \\
(10,90) & -0.64 & 0.05  & 0.12  & -1.92 & 0.06  & 0.22  & -1.67 & 0.07  & 0.14  & -8.57 & 0.12  & 0.38  & -2.51 & 0.07  & 0.14  & -16.39 & 0.25  & 0.66 \\\addlinespace
(15,30) & -9.84 & 0.17  & 0.99  & -22.38 & 0.78  & 4.05  & -14.18 & 0.29  & 1.18  & -31.72 & 2.15  & 5.09  & -14.50 & 0.29  & 1.18  & -32.49 & 3.28  & 6.13 \\
(15,45) & -7.48 & 0.12  & 0.33  & -18.60 & 0.34  & 1.01  & -12.02 & 0.17  & 0.42  & -30.73 & 1.92  & 5.59  & -12.80 & 0.17  & 0.42  & -32.44 & 3.16  & 6.04 \\
(15,60) & -4.82 & 0.09  & 0.17  & -12.73 & 0.24  & 0.66  & -9.84 & 0.17  & 0.50  & -28.41 & 1.00  & 5.81  & -11.00 & 0.17  & 0.50  & -31.65 & 3.37  & 7.56 \\
(15,75) & -3.22 & 0.10  & 0.20  & -9.99 & 0.19  & 0.55  & -6.98 & 0.12  & 0.28  & -23.74 & 0.41  & 3.63  & -8.02 & 0.12  & 0.28  & -28.63 & 3.08  & 8.39 \\
(15,90) & -2.07 & 0.07  & 0.13  & -6.33 & 0.12  & 0.38  & -4.86 & 0.12  & 0.25  & -18.58 & 0.29  & 0.58  & -6.08 & 0.12  & 0.25  & -25.48 & 2.61  & 9.02 \\\addlinespace
(20,30) & -11.78 & 0.46  & 4.81  & -26.11 & 2.09  & 8.49  & -15.34 & 1.08  & 3.58  & -33.11 & 4.49  & 7.99  & -15.55 & 1.08  & 3.58  & -33.65 & 5.44  & 9.76 \\
(20,45) & -9.11 & 0.17  & 0.47  & -21.80 & 1.13  & 6.78  & -13.85 & 0.57  & 4.81  & -33.09 & 4.86  & 16.42 & -14.52 & 0.57  & 4.81  & -34.40 & 6.55  & 16.95 \\
(20,60) & -7.71 & 0.18  & 0.66  & -19.10 & 0.74  & 4.52  & -12.68 & 0.29  & 0.82  & -32.32 & 3.57  & 9.36  & -13.33 & 0.29  & 0.82  & -34.14 & 6.23  & 12.20 \\
(20,75) & -5.15 & 0.15  & 0.30  & -13.87 & 0.75  & 4.37  & -10.16 & 0.22  & 0.66  & -29.27 & 2.30  & 9.15  & -11.03 & 0.22  & 0.66  & -32.26 & 5.76  & 15.64 \\
(20,90) & -3.98 & 0.14  & 0.37  & -11.00 & 0.30  & 0.56  & -8.03 & 0.23  & 0.45  & -25.50 & 1.47  & 6.16  & -9.17 & 0.23  & 0.45  & -30.44 & 5.09  & 16.37 \\\addlinespace
(25,30) & -11.69 & 0.83  & 4.99  & -25.91 & 3.46  & 12.47 & -14.92 & 1.73  & 7.21  & -32.17 & 6.35  & 18.22 & -15.19 & 1.73  & 7.21  & -32.73 & 7.74  & 13.89 \\
(25,45) & -10.87 & 0.32  & 2.69  & -24.38 & 2.55  & 15.51 & -15.11 & 0.74  & 2.64  & -33.76 & 7.01  & 14.90 & -15.58 & 0.74  & 2.64  & -34.65 & 9.75  & 20.71 \\
(25,60) & -9.18 & 0.25  & 0.51  & -21.57 & 2.13  & 8.09  & -14.12 & 0.45  & 1.82  & -33.82 & 6.46  & 20.65 & -14.75 & 0.45  & 1.82  & -35.23 & 9.11  & 26.48 \\
(25,75) & -7.32 & 0.23  & 0.48  & -18.44 & 1.69  & 5.01  & -12.50 & 0.35  & 0.92  & -32.56 & 8.07  & 27.64 & -13.34 & 0.35  & 0.92  & -34.55 & 8.80  & 19.91 \\
(25,90) & -6.22 & 0.22  & 0.41  & -15.82 & 0.71  & 3.38  & -11.02 & 0.35  & 0.81  & -30.33 & 4.66  & 26.77 & -12.15 & 0.35  & 0.81  & -33.83 & 10.51 & 26.92 \\\addlinespace
(30,30) & -12.51 & 2.18  & 8.97  & -26.66 & 7.53  & 29.84 & -15.54 & 3.31  & 11.85 & -32.30 & 11.95 & 34.28 & -15.73 & 3.31  & 11.85 & -32.81 & 13.30 & 34.73 \\
(30,45) & -11.26 & 0.41  & 2.07  & -25.20 & 3.93  & 29.24 & -15.55 & 1.08  & 7.44  & -34.21 & 11.48 & 28.04 & -15.81 & 1.08  & 7.44  & -34.80 & 14.30 & 37.37 \\
(30,60) & -10.57 & 0.43  & 1.24  & -24.26 & 3.36  & 10.33 & -15.50 & 1.08  & 4.66  & -35.18 & 9.84  & 24.00 & -15.85 & 1.08  & 4.66  & -36.01 & 16.53 & 49.18 \\
(30,75) & -8.36 & 0.43  & 1.55  & -20.66 & 4.33  & 14.04 & -13.94 & 0.75  & 2.88  & -33.98 & 8.92  & 28.41 & -14.65 & 0.75  & 2.88  & -35.54 & 11.87 & 26.38 \\
(30,90) & -7.62 & 0.27  & 0.59  & -18.57 & 2.80  & 7.31  & -12.78 & 0.55  & 1.73  & -32.59 & 12.17 & 50.15 & -13.68 & 0.55  & 1.73  & -35.16 & 18.77 & 35.48 \\
\midrule
\textbf{Avg.} & \textbf{-6.11} & \textbf{0.25} &  & \textbf{-14.73} & \textbf{1.32} &  & \textbf{-9.85} & \textbf{0.47} &  & \textbf{-25.32} & \textbf{3.70} &  & \textbf{-10.70} & \textbf{0.47} &  & \textbf{-28.38} & \textbf{5.66} & \\\bottomrule
\end{tabular}
    \end{footnotesize}
\rmfamily
	\caption{Results for the single-block SPRP with decoupling for different values of travel time factor $\beta$ and picker capacity $C$.}\label{tab:decouple}
\end{table}%
 \end{landscape}
}
Concerning the performance of our formulation, we observe the tendency that average and maximum runtimes increase if either the time required by the picker traveling alone decreases, i.e., $c_p$ decreases, or if the capacity $C$ increases. Nevertheless, all instances can be solved to optimality within very short runtimes of at most 50 seconds. The average runtimes over all instances of a certain combination of picker speed and capacity range between 0.25 and 5.66 seconds. We find that decoupling results in considerable cost savings, which grow with increasing picker speed and capacity. Even assuming the conservative values $C=2$ and $\beta=0.75$, the objective value is notably reduced by around 6\% on average. For the most optimistic setting with $C=6$ and $\beta=0.5$, savings rise to more than 28\%. 

\subsection{Results for the Single-Block SPRP with Multiple End Depots}
\label{sec:multdepot}
To study the effect of multiple end depots, we generate $3 \cdot 900 = 2700$ new instances from the standard SPRP instances of \citet{Scholz:2016} by locating end depots at the top or the bottom (independent of each other) of each aisle of the respective instance with a probability of $\sigma \in \{0.1, 0.5,1.0\}$. Thus, the instances with $\sigma=1$ refer to the single-block SPRP with decentralized depositing introduced in \citet{DeKoster:1998}. Table~\ref{tab:multEndDepot} presents aggregate results for each instance group. Columns $\Delta$ report the average gap between the optimal solution with multiple end depots and the optimum of the standard setting in which the picker must return to the start depot. Columns \runtime{avg} and \runtime{max} again report average and maximum runtime for each instance group.

Our formulation is able to solve all instances to optimality within a maximum runtime of approximately seven seconds. We note that, in general, the runtimes slightly increase with $\sigma$, on average from 1.13 seconds to 2.09 seconds. With respect to solution quality, we see that only moderate average savings between 2\% for $\sigma=0.1$ and 3.4\% for $\sigma=1.0$ can be realized. This suggests that, in a single-block rectangular warehouse, the overall benefits of multiple end depot are rather limited. This result might be different if the batching decision already incorporates the availability of multiple end depots. The results also indicate that already a few additional end depots achieve a large portion of the possible benefits.

\begin{table}[htbp]
  \centering
    \begin{footnotesize}
\sffamily
\ra{1.1}
\setlength{\tabcolsep}{3.0pt}
\begin{tabular}{@{}l@{\hspace{0.3cm}}rrr@{\hspace{0.3cm}}rrr@{\hspace{0.3cm}}rrr@{}}
\toprule
&  \multicolumn{3}{c}{$\sigma = 0.1$} & \multicolumn{3}{c}{$\sigma = 0.5$} & \multicolumn{3}{c}{$\sigma = 1.0$} \\ \cmidrule(l{.1cm}r{.1cm}){2-4} \cmidrule(l{.1cm}r{.1cm}){5-7} \cmidrule(l{.1cm}r{.0cm}){8-10}
$(\numberOfAisles,\numberOfArticles)$  & $\Delta$ & \runtime{avg} & \runtime{max}  & $\Delta$ & \runtime{avg} & \runtime{max}  & $\Delta$ & \runtime{avg} & \runtime{max} \\ \midrule
(5,30) & -1.25 & 0.02  & 0.05  & -2.78 & 0.03  & 0.09  & -3.24 & 0.03  & 0.08 \\
(5,45) & -1.68 & 0.03  & 0.09  & -3.82 & 0.03  & 0.08  & -4.22 & 0.03  & 0.09 \\
(5,60) & -3.21 & 0.03  & 0.09  & -6.85 & 0.02  & 0.05  & -7.26 & 0.02  & 0.06 \\
(5,75) & -3.75 & 0.04  & 0.12  & -6.47 & 0.03  & 0.13  & -7.48 & 0.03  & 0.05 \\
(5,90) & -3.87 & 0.04  & 0.12  & -7.46 & 0.03  & 0.05  & -7.67 & 0.03  & 0.05 \\\addlinespace
(10,30) & -1.25 & 0.11  & 0.38  & -2.51 & 0.26  & 0.51  & -2.83 & 0.28  & 0.53 \\
(10,45) & -1.55 & 0.11  & 0.31  & -2.50 & 0.17  & 0.46  & -2.76 & 0.21  & 0.42 \\
(10,60) & -1.31 & 0.10  & 0.25  & -2.39 & 0.17  & 0.47  & -2.55 & 0.17  & 0.36 \\
(10,75) & -1.27 & 0.08  & 0.16  & -2.47 & 0.13  & 0.45  & -2.59 & 0.18  & 0.58 \\
(10,90) & -1.86 & 0.08  & 0.19  & -3.06 & 0.12  & 0.30  & -3.22 & 0.11  & 0.36 \\\addlinespace
(15,30) & -1.96 & 0.32  & 1.53  & -2.71 & 0.70  & 3.27  & -2.97 & 0.80  & 2.62 \\
(15,45) & -1.32 & 0.29  & 0.75  & -2.45 & 0.73  & 2.90  & -2.58 & 0.98  & 3.68 \\
(15,60) & -1.65 & 0.29  & 0.93  & -2.40 & 0.59  & 2.14  & -2.48 & 0.76  & 3.54 \\
(15,75) & -1.68 & 0.22  & 0.60  & -2.56 & 0.30  & 0.79  & -2.68 & 0.39  & 1.36 \\
(15,90) & -1.98 & 0.19  & 0.63  & -2.68 & 0.20  & 0.39  & -2.74 & 0.22  & 0.53 \\\addlinespace
(20,30) & -2.53 & 0.75  & 2.70  & -3.42 & 1.69  & 4.55  & -3.57 & 2.10  & 4.94 \\
(20,45) & -2.27 & 0.70  & 2.66  & -2.92 & 1.73  & 3.81  & -3.08 & 1.98  & 5.08 \\
(20,60) & -1.45 & 1.18  & 3.15  & -2.33 & 1.81  & 4.31  & -2.37 & 2.68  & 6.07 \\
(20,75) & -1.49 & 0.76  & 3.37  & -2.16 & 1.51  & 4.73  & -2.27 & 1.69  & 4.97 \\
(20,90) & -1.74 & 0.72  & 1.96  & -2.52 & 1.12  & 3.96  & -2.65 & 1.30  & 5.47 \\\addlinespace
(25,30) & -3.23 & 1.31  & 4.36  & -4.31 & 2.95  & 6.46  & -4.48 & 3.77  & 7.49 \\
(25,45) & -2.44 & 1.97  & 5.48  & -3.06 & 3.57  & 9.39  & -3.24 & 4.10  & 8.78 \\
(25,60) & -1.77 & 2.09  & 5.88  & -2.44 & 3.81  & 8.29  & -2.60 & 3.62  & 7.23 \\
(25,75) & -1.79 & 1.73  & 6.34  & -2.40 & 3.26  & 8.60  & -2.53 & 3.08  & 7.40 \\
(25,90) & -2.02 & 1.56  & 5.23  & -2.63 & 2.43  & 5.67  & -2.69 & 3.06  & 9.80 \\\addlinespace
(30,30) & -3.16 & 3.65  & 8.22  & -4.20 & 5.01  & 11.12 & -4.43 & 5.50  & 12.12 \\
(30,45) & -2.53 & 3.74  & 8.93  & -3.04 & 5.54  & 12.90 & -3.19 & 6.63  & 17.93 \\
(30,60) & -2.13 & 4.30  & 9.40  & -2.55 & 5.92  & 10.22 & -2.62 & 7.05  & 11.18 \\
(30,75) & -2.09 & 3.67  & 10.50 & -2.48 & 5.32  & 13.70 & -2.62 & 5.90  & 14.82 \\
(30,90) & -1.77 & 3.77  & 12.93 & -2.16 & 5.36  & 14.52 & -2.24 & 5.89  & 14.15 \\
\midrule
\textbf{Avg.} & \textbf{-2.07} & \textbf{1.13} &  & \textbf{-3.19} & \textbf{1.82} &  & \textbf{-3.39} & \textbf{2.09} &  \\\bottomrule
\end{tabular}%
  \end{footnotesize}
\rmfamily
	\caption{Results for single-block SPRP with multiple end depots for different probabilites $\sigma$ that an aisle contains an end depot at the top or bottom.}
  \label{tab:multEndDepot}%
\end{table}%

\section{Conclusion}
\label{sec:conclusion}
In this paper, we present a compact formulation of the standard SPRP that directly exploits two properties of an optimal picking tour used in the algorithm of \citet{Ratliff:1983} and thus does not require classical subtour elimination constraints. Our formulation outperforms exisiting standard SPRP formulations from the literature and is able to solve large problem instances within short runtimes. The extensions of our formulation to scattered storage, decoupling of picker and cart, and multiple end depots are also able to solve realistically sized instances with low computational effort. Large savings are possible by allowing the decoupling of picker and cart: assuming a picker capacity of only two items and a reduction of travel time of $25\%$ when the picker travels alone, cost savings of $6\%$ are possible, and up to $28\%$ are achieved with a picker capacity of six items and a doubling of picker speed. Contrary to this, the cost savings of multiple end depots are rather limited. An interesting topic for future research is the extension or utilization of our formulation for integrated warehousing problems with a picker routing component, as outlined in Section 1.

\section*{Acknowledgements} We gratefully acknowledge funding from the AIF under grant number IGF-19570~N/2.
\newpage
\bibliography{bib}
\end{document}